
\input harvmac.tex

\overfullrule=0mm
\def\ra{\rangle}

\def\IR{\relax{\rm I\kern-.18em R}}
\font\cmss=cmss10 \font\cmsss=cmss10 at 7pt
\def\IZ{\relax\ifmmode\mathchoice
{\hbox{\cmss Z\kern-.4em Z}}{\hbox{\cmss Z\kern-.4em Z}}
{\lower.9pt\hbox{\cmsss Z\kern-.4em Z}}
{\lower1.2pt\hbox{\cmsss Z\kern-.4em Z}}\else{\cmss Z\kern-.4em Z}\fi}

\def\inbar{\,\vrule height1.5ex width.4pt depth0pt}
\def\IB{\relax{\rm I\kern-.18em B}}
\def\IC{\relax\hbox{$\inbar\kern-.3em{\rm C}$}}
\def\ID{\relax{\rm I\kern-.18em D}}
\def\IE{\relax{\rm I\kern-.18em E}}
\def\IF{\relax{\rm I\kern-.18em F}}
\def\IG{\relax\hbox{$\inbar\kern-.3em{\rm G}$}}
\def\IH{\relax{\rm I\kern-.18em H}}
\def\II{\relax{\rm I\kern-.18em I}}
\def\IK{\relax{\rm I\kern-.18em K}}
\def\IL{\relax{\rm I\kern-.18em L}}
\def\IM{\relax{\rm I\kern-.18em M}}
\def\IN{\relax{\rm I\kern-.18em N}}
\def\IO{\relax\hbox{$\inbar\kern-.3em{\rm O}$}}
\def\IP{\relax{\rm I\kern-.18em P}}
\def\IQ{\relax\hbox{$\inbar\kern-.3em{\rm Q}$}}
\def\IGa{\relax\hbox{${\rm I}\kern-.18em\Gamma$}}
\def\IPi{\relax\hbox{${\rm I}\kern-.18em\Pi$}}
\def\ITh{\relax\hbox{$\inbar\kern-.3em\Theta$}}
\def\IOm{\relax\hbox{$\inbar\kern-3.00pt\Omega$}}


\def\~{\tilde }
\def\^{\hat }
\def\={\bar }
\def\rank{{\rm rank }}

\def\lra{\leftrightarrow}
\def\CJ{{\cal J} }
\def\CE{{\cal E} }
\def\T{{\scriptscriptstyle T} }
\def\A{{\scriptscriptstyle A} }

\def\bbuildrel#1_#2^#3{\mathrel{\mathop{\kern 0pt#1}\limits_{#2}^{#3}}}
\def\bk{\bbuildrel{=}_{k}^{}  }

\catcode`\@=11
\def\displaylinesno#1{\displ@y\halign{
	\hbox to\displaywidth{$\@lign\hfil\displaystyle##\hfil$}&
	\llap{$##$}\crcr
#1\crcr}}
\catcode`\@=12

\def\sitle#1#2{\nopagenumbers\abstractfont\line{#1}%
\vskip .5in\centerline{\titlefont #2}\abstractfont\vskip .2in\pageno=0}
\def\meti#1{\par\indent\llap{#1\enspace}\ignorespaces}
\def\myfni{e-mail: mbhalpern@lbl.gov,
halpern@physics.berkeley.edu }
\def\myfnii{e-mail: sochen@asterix.lbl.gov }


\def\acknowledge{
This work was supported in part by the Director, Office of Energy
Research, Office of High Energy and Nuclear Physics, Division of High
Energy Physics of the U.S. Department of Energy under Contract
DE-AC03-76SF00098 and in part by the National Science Foundation
under
grant PHY-90-21139.
}



\line{June 1994\hfill }
\sitle{hep-th/9406076\hfill LBL-35718, UCB-PTH-94/15}{\vbox{
\centerline{Flat Connections for Characters}
\vskip .2cm
\centerline{ in Irrational Conformal Field Theory}
}}
\centerline{ M. B. Halpern\footnote{$^\star $}{\myfni} and
N. Sochen\footnote{$^*$}{\myfnii} }
\vskip0.4cm
\centerline{\it Department of Physics\footnote{$^\dagger$}{\acknowledge}}
\centerline{\it University of California at Berkeley }
\centerline{\it and}
\centerline{\it Theoretical Physics group}
\centerline{\it Lawrence Berkeley Laboratory}
\centerline{\it Berkeley, CA 94720, U.S.A.}
\vskip0.4cm
\centerline{\bf Abstract }
Following the paradigm on the sphere, we begin the study of irrational
conformal field theory (ICFT) on the torus. In particular, we find that
the affine-Virasoro characters of ICFT  satisfy heat-like differential
equations with flat connections. As a first example, we solve the system
for the general $g/h$ coset construction, obtaining
an integral representation for the general coset characters. In a second
application, we solve for the high-level characters of the general ICFT
on simple $g$, noting a simplification for the subspace of theories which
possess a non-trivial symmetry group. Finally, we give a geometric
formulation of the  system in which the flat connections are
generalized Laplacians on the centrally-extended loop group.

\vfill\eject
\newsec{Introduction}

\footline={\hss\tenrm\folio\hss}
Affine Lie algebra, or current algebra on $S^1$, was discovered independently
in mathematics
\ref\KM{V.G. Ka\v c, Anal. App. {\bf 1} (1967) 328\semi
R.V. Moody, Bull. Am. Math. Soc. {\bf 73} (1967) 217.}
and physics
\ref\BH{K. Bardak\c ci and M.B. Halpern, Phys. Rev. {\bf D3} (1971) 2493.}.
It is now understood that affine Lie algebra underlies both rational
conformal field theory (RCFT) and irrational conformal field theory (ICFT),
which includes RCFT as a small subspace,
\nref\HK{M.B. Halpern and E. Kiritsis, Mod. Phys. Lett. {\bf A4}
(1989) 1373\semi Erratum ibid. {\bf A4} (1989) 1797.}
\nref\Moroz{A. Yu Morozov, A.M. Perelomov, A.A. Rosly, M.A. Shifman and
A.V. Turbiner, Int. J. Mod. Phys. {\bf A5} (1990) 803.}
\eqn\Ii{
{\rm ICFT}\supset\supset {\rm RCFT}\quad .
}
At present, the only known path into the space of ICFT's
is the general affine-Virasoro construction whose conformal stress tensors
have the form \refs{\HK,\Moroz},
\nref\H{M.B. Halpern, Phys. Rev. {\bf D4} (1971) 2398.}
\nref\Sug{E. Witten, Commun. Math. Phys. {\bf 92} (1984) 455\semi
V.G. Knizhnik and A.B. Zamolodchikov, Nucl. Phys. {\bf B247} (1984) 83\semi
G. Segal, unpublished.}
\nref\GKO{P. Goddard, A. Kent and D. Olive, Phys. Lett. {\bf B152} (1985) 88.}
\eqn\Iii{
T(L)=L^{ab}{}^*_*J_aJ_b{}^*_*
}
where $J_a,\ a=1\dots {\rm dim\ }g$ are the currents of affine $g$ and
$L^{ab}$ is a solution of the Virasoro master equation \refs{\HK,\Moroz}. The
space of solutions of the master equation,
called affine-Virasoro space, includes the affine-Sugawara
constructions \refs{\BH,\H,\Sug},
the coset constructions \refs{\BH,\H,\GKO}
and a vast number
of unitary constructions with irrational central charge
\ref\five{M.B. Halpern, E.B. Kiritsis,  N.A. Obers, M. Porrati and J.P. Yamron,
Int. J. Mod. Phys. {\bf A5} (1990) 2275.}.
See Reference
\ref\rev{M.B. Halpern, {\it ``Recent Developments in the Virasoro Master
Equation''}, in the proceedings of the Stony Brook conference, {\it Strings
and Symmetries} 1991, World Scientific, 1992.}
for a brief survey of affine-Virasoro space and its
partial classification by graph theory.

One of the most prominent features of
affine-Virasoro space is K-conjugation covariance \refs{\BH,\H,\GKO,\HK}, which
says that affine-Virasoro constructions come in commuting K-conjugate
pairs, $T=T(L)$ and
$\~T=T(\~L)$, which sum to the affine-Sugawara
construction $T_g$ on $g$,
\nref\Hen{M.B Halpern, Ann. of Phys. {\bf 194} (1989) 247.}
\nref\obersI{M.B Halpern and N.A. Obers, Int. J. Mod. Phys.
{\bf A9} (1994) 265.}
\eqn\Iiib{
\~T+T=T_g\quad ,\quad \~c+c=c_g\quad .
}
Thus, each K-conjugate pair of ICFT's naturally forms a biconformal field
theory \refs{\Hen,\obersI},
complete with biprimary and bisecondary
fields, whose biconformal correlators must be factorized to obtain the
conformal correlators of the individual ICFT's.

\nref\obersII{M.B. Halpern and N.A. Obers, Int. J. Mod. Phys.
{\bf A9} (1994) 419.}
Recently, dynamical equations for the biconformal correlators, the
affine-Virasoro Ward identities \refs{\obersI,\obersII},
have been obtained for ICFT on the sphere. These may be expressed as
generalized
Knizhnik-Zamolodchikov equations
\ref\obersIII{M.B. Halpern and N.A. Obers, {\it Flat Connections and Non-Local
Conserved Quantities in Irrational Conformal Field Theory} Berkeley/Ecole
Polytechnique preprint UCB-PTH-93/33, CPTH-A277.1293, hep-th/9312050},
\eqna\Iiii
$$
\displaylinesno{
\=\del_iA(\=z,z)=A(\=z,z)\overline{W}_i(\=z,z)\quad ,\quad
\del_iA(\=z,z)=A(\=z,z)W_i(\=z,z)&\Iiii a\cr
A_g(z)=A(z,z)&\Iiii b\cr}
$$
for the biconformal correlators $A$. The affine-Virasoro connections
$\overline{W}$,
 $W$ are flat connections and the affine-Sugawara correlators $A_g$ are
obtained from the biconformal correlators when $\=z=z$. These equations
have been solved for the correlators of the general $g/h$ coset construction
\refs{\obersI,\obersII,\obersIII},
providing a first-principle derivation of the coset blocks first proposed
by Douglas
\ref\Doug{M.R. Douglas, {\it ``G/H Conformal Field Theory''}, Caltech preprint,
CALT-68-1453, 1987, unpublished.}.
Moreover, the equations have been solved for the
general ICFT at high level
\refs{\obersII,\obersIII}
on simple $g$, although the resulting
high-level conformal correlators have not yet been analyzed at the level of
conformal blocks. See Ref. \ref\Hnt{M.B. Halpern, {\it ``Recent Progress in
Irrational Conformal Field Theory''}, Berkeley preprint, UCB-PTH-93/25,
hep-th/9309087, 1993. To appear in the proceedings of the Berkeley conference,
{\it Strings 1993}.}
for a review of the affine-Virasoro Ward identities, including the general
solution \obersII\ which exhibits braiding for all ICFT.

In this paper, we begin the study of ICFT on the torus, following the paradigm
on the sphere. In particular, we study the affine-Virasoro characters
(the bicharacters),
\eqn\Ivi{
\chi(T,\~\tau,\tau,g)=Tr_{\T}\left(\~q^{\~L(0)-\~c/24}q^{L(0)-c/24}g\right)
}
where $\~L(0)$ and $L(0)$ are the zero modes  of a K-conjugate pair of
stress tensors on affine $g$. Here, the trace  is over the  affine irrep $V_T$,
the source $g\in G$ is an element of the Lie group, and, as on the sphere,
the
affine-Sugawara (or affine) characters are obtained from the bicharacters
at $\~\tau=\tau$,
\eqn\Ivii{
\chi_g(T,\tau,g)=
\chi(T,\tau,\tau,g)=Tr_{\T}\left(q^{L_g(0)-c_g/24}g\right)\quad .
}
Given \Ivii\ as a boundary condition,
we find that the bicharacters are the unique solutions of heat-like
differential equations with flat connections $\~D$ and $D$,
\eqna\Iviii
$$
\eqalignno{
\del_{\~\tau}\chi(T,\~\tau,\tau,g)&=
\~D(\~\tau,\tau,g)\chi(T,\~\tau,\tau,g)\quad &\Iviii a\cr
\del_\tau\chi(T,\~\tau,\tau,g)&=
D(\~\tau,\tau,g)\chi(T,\~\tau,\tau,g)\quad &\Iviii b\cr}
$$
whose existence and properties are the central subject of this paper.
These systems include and generalize the heat equations for the
affine-Sugawara characters obtained by  Bernard
\ref\Bernard{D. Bernard, Nucl. Phys. {\bf B303} (1988) 77.}
and by Eguchi and Ooguri
\ref\OogI{T. Eguchi and H. Ooguri, Nucl. Phys. {\bf B313} (1989) 492.}.

Following the development on the sphere, we solve the system
first for the simple case of $h$ and the $g/h$ coset constructions, obtaining
a new integral representation for the general coset characters.

In a second application, we solve for the high-level form of the general
bicharacters
on simple $g$, noting a simplification on the subspace of $H$-invariant
CFT's \ref\Lieh{M.B. Halpern, E. Kiritsis and N.A. Obers, {\it The
Lie h-Invariant Conformal Field Theories and the Lie h-invariant Graphs},
in Infinite Analysis, Part A, Advanced Series in Mathematical Physics
{\bf 16}, Proceedings of the RIMS Project, World Scientific, 1992},
\eqn\Iix{
\hbox{ {\rm ICFT} }\supset\supset \hbox{ {\rm $H$-invariant\ CFT's} }
\supset\supset \hbox{ {\rm Lie $h$-invariant CFT's} }\supset\supset
\hbox{ {\rm RCFT } }\ .
}
The set of $H$-invariant CFT's is the subset of all ICFT's which possess
a residual global symmetry group $H$,
which may be a finite group or a Lie group. Those theories with a Lie
invariance are
called the Lie $h$-invariant CFT's \Lieh, which include the affine-Sugawara
constructions and the coset constructions as a small subspace.

The simplification occurs when the source is chosen in the symmetry group
of the theory, and, as seen in the hierarchy \Iix, this simplification
occurs for the coset constructions as well.  Using intuition gained from the
cosets, we factorize the high-level bicharacters of the Lie $h$-invariant CFT's
to obtain a set of high-level candidate characters for this class of
ICFT's. The set of candidate characters correctly includes the high-level
form of the coset characters and should be further analyzed with respect to
modular covariance in the general case.

Finally we give a geometric formulation of the system on an affine source,
where the flat connections are generalized Laplacians on the
centrally-extended loop group. These Laplacians involve new first-order
differential representations of affine Lie algebra.

\newsec{The General Affine-Virasoro Construction}

In this Section, we review some aspects of ICFT which will be relevant in
the development below.

The general affine-Virasoro construction
begins with the currents $J_a$ of untwisted affine Lie $g$ [1,2],
\eqna\comrelg
$$
\displaylinesno{
[J_a(m),J_b(n)]=i f_{ab}^{\ \ c}J_c(m+n)+mG_{ab}\delta_{m+n,0}&\comrelg a\cr
a,b=1\dots {\rm dim} g\quad  ,\quad   m,n\in Z&\comrelg b\cr}
$$
where $f_{ab}^{\ \ c}$ and $G_{ab}$ are respectively the structure constants
and
general Killing metric of $g=\oplus_Ig_I$. To obtain invariant levels
$x_I=2k_I/\psi_I^2$
of $g_I$ with dual Coxeter numbers $\~h_I=Q_I/\psi_I^2$, take
\eqn\GandF{
G_{ab}=\oplus_Ik_I\eta_{ab}^I\quad ,\quad f_{ac}^{\ \ d}
f_{bd}^{\ \ c}=-\oplus_IQ_I\eta_{ab}^I
}
where $\eta_{ab}^I$ and $\psi_I$ are respectively the Killing metric and
highest root of $g_I$.

The stress tensors of the general affine-Virasoro construction are
elements of the enveloping algebra of the affine algebra
\refs{\HK,\Moroz},
\eqn\stress{
T(L)=L^{ab}{}^*_*J_aJ_b{}^*_*=\sum_{m\in \IZ}L(m)z^{-m-2}
}
where $L^{ab}=L^{ba}$ is called the inverse inertia tensor in analogy with the
spinning top. In order that the modes $L(m)$ generate the Virasoro algebra,
\eqn\Vir{
[L(m),L(n)]=(m-n)L(m+n)+{c\over 12}m(m^2-1)\delta_{m+n,0}
}
the inverse inertia tensor must satisfy the Virasoro master equation
\refs{\HK,\Moroz},
\eqna\mastereq
$$
\displaylinesno{
L^{ab}=2L^{ac}G_{cd}L^{db}-L^{cd}L^{ef}f_{ce}^{\ \ a}f_{df}^{\ \ b}-
L^{cd}f_{ce}^{\ \ f}f_{df}^{\ \ (a}L^{b)e}&\mastereq a\cr
c(L)=2G_{ab}L^{ab}&\mastereq b\cr}
$$
where $c(L)$ is the central charge of the CFT. The master equation has been
identified
\ref\HY{M.B. Halpern and J.P. Yamron, Nucl. Phys. {\bf B332} (1990) 411.}
as an Einstein-like system on the
group manifold with central charge $c=\dim g-4R$, where $R$ is the
Einstein curvature scalar.

\meti{A.}Affine-Sugawara constructions
\refs{\BH,\H,\Sug}.
The affine-Sugawara construction $L_g$ is
\eqn\Lg{
L_g^{ab}=\oplus_I{\eta^{ab}_I\over 2k_I+Q_I}\quad ,\quad
c_g=\sum_I{x_I\dim g_I\over x_I+\~h_I}
}
for arbitrary levels of affine $g$, and similarly for $L_h$ when
$h\subset g$.
\meti{B.}K-conjugation covariance
\refs{\BH,\H,\GKO,\HK}. When $L$ is a solution of the master
equation on $g$, then so is the K-conjugate partner of $L$, called $\~L$,
\eqn\Ltwidle{
\~L^{ab}=L_g^{ab}-L^{ab}\quad ,\quad c(\~L)=c_g-c(L)
}
while the corresponding stress tensors $T\equiv T(L)$ and $\~T\equiv T(\~L)$
form a commuting pair of Virasoro operators which sum to the affine-Sugawara
construction,
\eqn\Tg{
\~T+T=T_g\quad ,\quad\~c+c=c_g\quad ,\quad T(z)\~T(w)=regular\quad .
}
As the simplest examples of K-conjugation, the $g/h$ coset constructions
$\~T=T_{g/h}=T_g-T_h$
\refs{\BH,\H,\GKO}
are the K-conjugate partners of $T_h$ on $g$.

More generally, each breakup $T_g=\~T+T$ naturally defines a biconformal
field theory \refs{\Hen,\obersI}, with two commuting Virasoro operators.
The breakup also suggests that the affine-Sugawara construction is a tensor
product CFT, formed by tensoring the conformal theories of $\~T$ and $T$.
In practice, we face the inverse problem, namely to factorize
\refs{\Hen,\obersI,\obersII,\obersIII} the affine-Sugawara blocks into the
conformal blocks of $\~T$ and $T$.
\meti{C.}$T,J$ commutator. The commutator of the stress tensor
with the currents is \HK,
\nref\HO{M.B. Halpern and N.A. Obers, Nucl. Phys. {\bf B345} (1990) 607.}
\nref\HYII{M.B. Halpern and J.P. Yamron, Nucl. Phys. {\bf B351} (1991) 333.}
\eqna\LJ
$$
\displaylinesno{
[L(m),J_a(n)]=
-nM(L)_a^{\ b}J_b(m+n)+N(L)_a^{\ bc}({}^*_*J_bJ_c{}^*_*)_{m+n}&\LJ a\cr
M(L)_a^{\ b}=2G_{ac}L^{cb}+f_{ad}^{\ \ e}L^{dc}f_{ce}^{\ \ b}\quad ,\quad
N(L)_a^{\ bc}=-if_{ad}^{\ (b}L^{c)d}&\LJ b\cr}
$$
and similarly for the K-conjugate theory with the substitution $L(m)\to\~L(m)$,
$L^{ab}\to\~L^{ab}$.
\meti{D.}High-level solutions. At high-level on simple $g$, the high-level
smooth solutions of the master equation have the form \refs{\HO,\HYII},
\eqna\hkL
$$
\displaylinesno{
\~L^{ab}={1\over 2k}\eta^{ac}\~P_c^{\ b}+O(k^{-2})\quad ,\quad
L^{ab}={1\over 2k}\eta^{ac}P_c^{\ b}+O(k^{-2})\quad &\hkL a\cr
\~L^{ab}+L^{ab}=L_g^{ab}={1\over 2k}\eta^{ab}+O(k^{-2})\quad &\hkL b\cr
\~c=\rank\~P+O(k^{-1})\quad ,\quad c=\rank P+O(k^{-1})\quad ,
\quad c_g=\dim g+O(k^{-1})&\hkL c\cr}
$$
where $\~P$ and $P$ are the high-level projectors of the $\~L$ and $L$
theories respectively,
\eqn\projs{
\~P^2=\~P\quad ,\quad P^2=P\quad ,\quad \~P+P=1\quad ,\quad \~P P=P\~P=0\quad .
}
\nref\graph{M.B. Halpern and N.A. Obers, Commun. Math. Phys. {\bf 138} (1991)
63.}
In the partial classification of ICFT by graph theory \refs{\graph,\rev},
the projectors are the adjacency matrices of the graphs, each of which labels
a level family of ICFT's.
\meti{E.}Symmetry groups in ICFT \Lieh. The generic ICFT on $g$ has no residual
global symmetry \graph,
but it is useful to distinguish the subspace of $H$-invariant
CFT's, which are all ICFT's with a residual global symmetry $H$,
\eqn\IIix{
\hbox{ {\rm ICFT} }\supset\supset\hbox{ {\rm $H$-invariant\ CFT's} }
\supset\supset\hbox{\rm Lie $h$-invariant CFT's }
\supset\supset\hbox{ {\rm RCFT } }\ .
}
The $H$-invariant CFT's on $g$ satisfy
\eqn\Hinv{
\~L=w\~Lw^{-1}\quad ,\quad L=wLw^{-1}\quad ,\quad
\forall w\in H\subset {\rm Aut} G
}
where $G$ is  the Lie group whose algebra is $g$. If $H$ is also a Lie
group, with Lie algebra $h\subset g$, we have the proper subspace of Lie
$h$-invariant CFT's, which satisfy
\eqn\TadjL{
[T_A^{\rm adj},\~L]= [T_A^{\rm adj},L]=0\quad ,\quad A=1,\ldots,\dim h
}
where $T^{\rm adj}_a$, $a=1,\ldots ,\dim g$ is the adjoint representation
of $g$. As seen in the
hierarchy \IIix, the Lie $h$-invariant CFT's include $h$ and the $g/h$
coset constructions as a small subspace.

\newsec{The Affine-Virasoro Characters}

For each K-conjugate pair $\~T$, $T$ of affine-Virasoro constructions on
compact
$g$, the affine-Virasoro (or biconformal) characters are defined as,
\eqn\bichar{
\chi(T,\~\tau,\tau,h)=Tr_{\T}\left(\~q^{\~L(0)-\~c/24}q^{L(0)-c/24}h\right)
}
where $q=e^{2\pi i\tau}$ ($\~q=e^{2\pi i\~\tau}$) with Im$\tau>0$
(Im$\~\tau >0$), and
\eqna\AV
$$
\eqalignno{
\~L(0)&=\~L^{ab}\big(J_a(0)J_b(0)+2\sum_{n>0}J_a(-n)J_b(n)\big)&\AV a\cr
L(0)&=L^{ab}\big(J_a(0)J_b(0)+2\sum_{n>0}J_a(-n)J_b(n)\big)&\AV b \cr}
$$
are the zero modes of $\~T$ and $T$. For flexibility below, we specify
the source $h$ in \bichar\ to be an element of the compact Lie group
$H\subset G$, which may be parametrized, for example, as
\eqn\hlocal{
h=\e{i\beta^A(x)J_A(0)}\quad ,\quad A=1,\ldots ,\dim h\quad
}
where $x^i$, $i=1,\ldots ,\dim h$ are coordinates on the $H$ manifold.
As special cases, we may then choose, if desired, the standard  sources
on $G$ or Cartan $G$ employed in Refs. \Bernard\ and
\OogI\ respectively.

In \bichar, the trace is over the integrable affine irrep $V_T$ whose affine
primary states $|R_T\ra$ correspond to matrix irrep $T$ of $g$.
In an $L$-basis of $T$
\refs{\Hen,\five,\obersI},
these are
called the $L^{ab}$-broken affine primary states, which satisfy,
\eqna\brokenbasis
$$
\displaylinesno{
J_a(m)|R_T\rangle^\alpha=\delta_{m,0}|R_T\rangle^\beta(T_a)_\beta^{\ \alpha}
\quad ,m\ge 0&\brokenbasis a\cr
\~L^{ab}(T_aT_b)_\alpha^{\ \beta}=\~\Delta_\alpha(T)\delta_\alpha^\beta\quad
,\quad L^{ab}(T_aT_b)_\alpha^{\ \beta}=
\Delta_\alpha(T)\delta_\alpha^\beta &\brokenbasis b\cr
\~L(0)|R_T\rangle^\alpha=\~\Delta_\alpha(T)|R_T\rangle^{\alpha}\quad ,\quad
L(0)|R_T\rangle^\alpha=\Delta_\alpha(T)|R_T\rangle^{\alpha}&\brokenbasis c\cr
\~\Delta_\alpha(T)+ \Delta_\alpha(T)= \Delta_g(T)&\brokenbasis d\cr}
$$
where $\~\Delta_\alpha(T)$, $\Delta_\alpha(T)$ and $\Delta_g(T)$ are the
conformal weights of the broken affine primaries under $\~T$, $T$ and $T_g$
respectively.
More generally, $L$-bases are the eigenbases of the conformal weight
matrices, such as \brokenbasis{b}, which occur at each level of
the irrep. In what follows, we often refer to the affine-Virasoro
characters as the bicharacters.

In this Section, we confine our remarks to some simple properties of the
bicharacters.
\meti{1.}N-point correlators. The bicharacters \bichar\ are only
the simplest (zero point) correlators on the torus. Although we will not pursue
this here, N-point correlators on the torus can be defined, as on the sphere,
by insertion of biconformal fields
\refs{\Hen,\obersI}
in the trace.
\meti{2.} K-conjugation invariance. It is clear on inspection  that the
bicharacters satisfy the K-conjugation invariance,
\eqn\Kconjsym{
\chi(T,\~\tau,\tau,h)|_{L\lra\~L\atop\tau\lra\~\tau}=
\chi(T,\~\tau,\tau,h)
}
under exchange of the K-conjugate CFT's.
\meti{3.} Affine-Sugawara boundary condition.  Since $\~T+T=T_g$ and
$\~c+c=c_g$,
the affine-Virasoro characters reduce to the affine-Sugawara (or affine)
characters,
\eqn\AS{
\chi_g(T,\tau,h)=
\chi(T,\tau,\tau,h)=Tr_{\T}\left(q^{L_g(0)-c_g/24}h\right)
}
on the affine-Sugawara line $\~\tau=\tau$.
\eqna\smallqid
\eqna\smallq
\meti{4.} Small $\~q$ and $q$. In order to obtain the leading
terms of the bicharacters when $\~q$ and $q$ are small, we need the identities,
$$
\displaylinesno{
h|R_T\rangle^\alpha=|R_T\rangle^\beta h(T)_\beta^{\ \alpha}&\smallqid a\cr
hJ_a(n)h^{-1}=\Omega(h)_a^{\ b}J_b(n)\quad ,\quad
\Omega(h)=h(T^{adj})^{-1}&\smallqid b\cr}
$$
where $h(T)$ is the corresponding element of $H\subset G$ in matrix irrep
$T$ of $g$.  Then, with  \brokenbasis{}\ and the affine algebra \comrelg{},
we may easily compute the contributions of the lowest states,
$$
\eqalignno{
\chi(T,\~\tau,\tau,h)&=\sum_{\alpha=1}^{\dim T}
\~q^{\~\Delta_\alpha(T)-\~c/24}q^{\Delta_\alpha(T)-c/24}
h(T)_\alpha^{\ \alpha}+\cdots &\smallq a\cr
\chi(T=0,\~\tau,\tau,h)&=1+\sum_{A=1}^{\dim g}
\~q^{\~\Delta_A-\~c/24}q^{\Delta_A-c/24}
h(T^{adj})_\A^{\ \A}+\cdots\ .&\smallq b\cr}
$$
For the vacuum
bicharacter in \smallq{b}, the computation of the non-leading terms was
performed in the L-basis
$J_A(-1)|0\rangle$ of the one-current states, so that $\~\Delta_A$ and
$\Delta_A$ (with $\~\Delta_A+\Delta_A=\Delta_g=1$) are the conformal weights
of these states under $\~T$ and $T$.

\newsec{The Affine-Virasoro Ward Identities}

\subsec{Statement and Strategy}

In this Section, we establish and study the affine-Virasoro Ward
identities for the bicharacters, which have the form,
\eqna\AVWI
$$
\eqalignno{
\~\del^q\del^p\chi(T,\~\tau,\tau,h)\vert_{\~\tau=\tau}&=
D_{qp}(\tau,h)\chi_g(T,\tau,h)&\AVWI a\cr
\del\equiv\del_{\tau}=2\pi iq\del_q \quad, &\quad
\~\del\equiv\del_{\~\tau}=2\pi i\~q\del_{\~q }&\AVWI b\cr}
$$
where $\chi_g$ is the affine-Sugawara character given in eq. \AS.

The $h$-differential operators $D_{qp}(\tau,h)$, called the affine-Virasoro
connection moments, are defined by the formula,
\nref\OogII{T. Eguchi and H. Ooguri, Nucl. Phys. {\bf B282} (1987) 308.}
\eqn\hmoments{
D_{qp}(\tau,h)\chi_g(T,\tau,h)=
(2\pi i)^{q+p}Tr_{\T}\left(q^{L_g(0)-c_g/24}
(\~L(0)-\~c/24)^q(L(0)-c/24)^ph\right)
}
where the zero modes $L(0)$, $\~L(0)$ of the K-conjugate stress tensors
are given in eq. \AV{}.
Note that the quantities on the right side of \hmoments\
are averages in the affine-Sugawara theory, so the connection moments may be
computed in principle by the methods of Refs.
\refs{\OogII,\Bernard}.

Our strategy to obtain these results is as follows: According to the definition
\bichar, the
left side of \AVWI{}\ is equal to  the right side of \hmoments, so the Ward
identities
\AVWI{}\ follow if the right side of \hmoments\ is proportional to $\chi_g$.
This will emerge in the following method for the computation of the
connection moments.

\subsec{Proof by Computational Scheme}

The following inductive algebraic scheme for the computation of affine-Sugawara
averages
is equivalent to the methods of Refs.
\refs{\OogII,\Bernard}.

We organize the problem in terms of the basic quantity,
\eqn\BI{
Tr_{\T}\left(q^{L_g(0)}J_a(-n){\cal O}h\right)\quad ,\quad n\in\IZ
}
where $\cal O$ is any vector in the enveloping algebra, and,
for simplicity, we restrict the source $h$ to those subgroups
for which $G/H$ is a reductive coset space. In this case, we may choose
a basis $a=1,\ldots ,\dim g=(A,I)$, in which
\eqna\RSC
$$
\displaylinesno{
f_{AI}^{\ \ B}=G_{AI}=0&\RSC a\cr
A=1,\ldots ,\dim h\quad ,\quad I=1,\ldots ,\dim g/h\quad .&\RSC b\cr}
$$
Then we have the relations,
\eqna\FII
$$
\displaylinesno{
q^{L_g(0)}J_a(-n)=q^nJ_a(-n)q^{L_g(0)}&\FII a\cr
\Omega(h)_a^{\ b}=\pmatrix{
                       \rho(h)_A^{\ B} & 0 \cr
                          0  &\sigma(h)_I^{\ J} \cr}&\FII b\cr
hJ_A(-n)=\rho(h)_A^{\ B}J_B(-n)h\quad ,\quad
hJ_I(-n)=\sigma(h)_I^{\ J}J_J(-n)h &\FII c\cr }
$$
from \smallqid{}\ and \comrelg{}.
With these relations and cyclicity of the trace, the current in \BI\ can be
moved  first to the left, then to the right of the trace and finally to
the left of the source. Rewriting ${\cal O}J_a$ as the original product
$J_a\cal O$ plus the commutator, we may solve the relation for the original
quantities \BI\ to obtain the basic identities,
\eqna\BII
$$
\eqalignno{
Tr_{\T}\left(q^{L_g(0)}J_A(-n){\cal O}h\right)&=
\big({q^n\rho(h)\over 1-q^n\rho(h)}\big)_A^{\ \ B}
Tr_{\T}\left(q^{L_g(0)}\big[{\cal O},J_B(-n)\big]h\right)
\  ,n\ne 0\quad\quad \ &\BII a\cr
Tr_{\T}\left(q^{L_g(0)}J_I(-n){\cal O}h\right)&=
\big({q^n\sigma(h)\over 1-q^n\sigma(h)}\big)_I^{\ \ J}
Tr_{\T}\left(q^{L_g(0)}\big[{\cal O},J_J(-n)\big]h\right)
\  ,n\in \IZ\ .\quad\quad &\BII b\cr}
$$
The identity \BII{a}\ does not hold for $n=0$ since $1-\rho(h)$ is not
invertible. With these relations and the affine algebra we may iteratively
reduce the number of currents in the trace by one, except for the zero modes
$J_A(0)$ of the $h$-currents.

To include the zero modes of the $h$ current, we introduce
$h$-vielbeins $\=e_i^{\ A}$ and $e_i^{\ A}$,
\eqn\viel{
\=e_i(h)=-ih\del_i h^{-1}=\=e_i^{\ A}(h)J_A(0)\quad ,\quad
e_i(h)=-ih^{-1}\del_i h=e_i^{\ A}(h)J_A(0)
}
where $i,A=1,\ldots ,\dim h$, and Lie derivatives on $h$,
\eqna\LD
\eqna\LDh
$$
\displaylinesno{
{\=E}_A=-i\=e_A^{\ \ i}\del_i \quad ,\quad
E_A=-ie_A^{\ \ i}\del_i &\LD a\cr
[\=E_A,\=E_B]=if_{AB}^{\ \ \ C}\=E_C\quad ,\quad
[E_A,E_B]=if_{AB}^{\ \ \ C}E_C\quad ,\quad
[\=E_A,E_B]=0 &\LD b\cr}
$$
where $\=e_A^{\ \ i}$ and $e_A^{\ \ i}$ are the inverse $h$-vielbeins.
{}From the definitions in \LD{a}\ we find that
$$
\displaylinesno{
\=E_A(h)h=-J_A(0)h\quad ,\quad E_A(h)h=hJ_A(0)&\LDh a\cr
\=E_A(h)h(T)=-T_Ah(T)\quad ,\quad E_A(h)h(T)=h(T)T_A&\LDh b\cr}
$$
and, using \FII{}, we obtain the basic identity for the zero modes
of the h-currents,
\eqn\ZMII{
Tr_{\T}\left(q^{L_g(0)}J_A(0){\cal O}h\right)=
E_A(h)Tr_{\T}\left(q^{L_g(0)}{\cal O}h\right)\quad .
}
Taken together, the relations \BII{}\ and \ZMII\ allow a reduction by one
in the number of currents in any affine-Sugawara average.

Iterating this step, the averages on the right side of \hmoments\
may be reduced to differential operators on the one-current averages,
\eqna\ZMIII
$$
\displaylinesno{
Tr_{\T}\left(q^{L_g(0)-c_g/24}J_A(0)h\right)=E_A(h)\chi_g(T,\tau,h)&\ZMIII a\cr
Tr_{\T}\left(q^{L_g(0)-c_g/24}J_I(0)h\right)=0&\ZMIII b\cr}
$$
which are proportional to the affine-Sugawara characters.
This completes the proof of the affine-Virasoro Ward identities \AVWI{}.

As an example, we have computed the first moment $D_{01}$ of the $L$ theory,
using
\eqna\DOlI
$$
\displaylinesno{
D_{01}(\tau,h)\chi_g(T,\tau,h)
=2\pi iTr_{\T}\left(q^{L_g(0)-c_g/24}(L(0)-c/24)h\right)\hfill &\DOlI a\cr
L(0)=L^{ab}J_a(0)J_b(0)
+2L^{ab}\sum_{n>0}J_a(-n)J_b(n) \hfill &\DOlI b\cr
\quad =L^{AB}J_A(0)J_B(0)+L^{AI}\big(J_A(0)J_I(0)+J_I(0)J_A(0)\big)
+L^{IJ}J_I(0)J_J(0) \hfill &\DOlI c\cr
\quad\ +2\sum_{n>0}
\left(L^{AB}J_A(-n)J_B(n)+L^{AI}\big(J_A(-n)J_I(n)+J_I(-n)J_A(n)\big)
+L^{IJ}J_I(-n)J_J(n)\right)\hfill &\cr}
$$
and the identities (4.6--11). The result is
\eqna\DOlII
$$
\eqalignno{
D_{01}(L,\tau,h)=2\pi i\{ -c/24 &+ L^{AB}E_A(h)E_B(h)+
L^{IJ}({\sigma(h)\over 1-\sigma(h)})_I^{\ K}(if_{JK}^{\ \ \ A}E_A(h)) & \cr
&+2L^{AB}\sum_{n>0}
({q^n\rho(h)\over 1-q^n\rho(h)})_A^{\ \ C}(if_{BC}^{\ \ \ D}E_D(h)+nG_{BC})&
\cr
&+2L^{IJ}\sum_{n>0}
({q^n\sigma(h)\over 1-q^n\sigma(h)})_I^{\ K}(if_{JK}^{\ \ \ A}E_A(h)+nG_{JK})\}
\ .  &\DOlII {} \cr}
$$
Similarly, the result for
$D_{10}$ is obtained  from \DOlII{}\ by the substitution
$L\to\~L$ and $c\to\~c$.

\subsec{The Affine-Sugawara Characters}

Adding the (1,0) and (0,1) Ward identities, we find the heat equation for
the affine-Sugawara characters,
\eqna\heatg
\eject
$$
\displaylinesno{
\del\chi_g(T,\tau,h)=D_g(\tau,h)\chi_g(T,\tau,h) &\heatg a\cr
D_g(\tau,h)=D_{01}+D_{10}\hfill & \cr
\hfill
=2\pi i\big\{ -c_g/24 + L_g^{AB}E_A(h)E_B(h)+
L_g^{IJ}({\sigma(h)\over 1-\sigma(h)})_I^{\ K}(if_{JK}^{\ \ \ A}E_A(h))
\quad\quad\quad\quad\quad & \cr
\hfill +2L_g^{AB}\sum_{n>0}
({q^n\rho(h)\over 1-q^n\rho(h)})_A^{\ \ C}(if_{BC}^{\ \ \ D}E_D(h)+nG_{BC})
\quad\quad\quad\quad\quad & \cr
\hfill +2L_g^{IJ}\sum_{n>0}
({q^n\sigma(h)\over 1-q^n\sigma(h)})_I^{\ K}(if_{JK}^{\ \ \ A}E_A(h)+nG_{JK})
\big\}
\ .\quad\quad\quad\quad &\heatg b \cr}
$$
The affine-Sugawara characters can also be understood as the simplest examples
of affine-Virasoro characters, obtained by choosing the simplest K-conjugate
pair $\~L=0$ and $L=L_g$. The heat-like equations in this case,
\eqna\heatgB
$$
\displaylinesno{
\del\chi(T,\~\tau,\tau,h)=D_g(\tau,h)\chi(T,\~\tau,\tau,h)
\quad ,\quad \~\del\chi(T,\~\tau,\tau,h)=0 &\heatgB a\cr
\chi(T,\~\tau,\tau,h)=\chi(T,\tau,\tau,h)=
\chi_g(T,\tau,h)\quad .&\heatgB b\cr}
$$
are equivalent to \heatg{}.
\eqna\heatBER

When the subgroup $H\subset G$ is taken to be $G$ itself, we recover
Bernard's result \Bernard\ on a $G$-source,
$$
\displaylinesno{
\del\chi_g(T,\tau,g)=D_g(\tau,g)\chi_g(T,\tau,g) &\heatBER a\cr
D_g(\tau,g)= 2\pi i\{ -c_g/24 + L^{ab}_gE_a(g)E_b(g)\hfill & \cr
\hfill {}+2L^{ab}_g\sum_{n>0}
({q^n\Omega(g)\over 1-q^n\Omega(g)})_a^{\ c}
(if_{bc}^{\ \ d}E_d(g)+nG_{bc})\}\quad\quad\quad\quad\quad &\heatBER b \cr}
$$
where $L_g^{ab}$ is given in eq. \Lg. For simple $g$,
Bernard also gives the alternate form of the $g$-connection,
\eqna\gcon
$$
\displaylinesno{
D_g(\tau,g)={2\pi i\over 2k+Q_g}\big\{ -(2k+Q_g)c_g/24 +
\eta^{ab}E_a(g)E_b(g) \hfill & \cr
\hfill {} + 2\eta^{ab}\big(E_a(g)\log \Pi(\tau,\Omega(g))\big)
E_b(g) -2kq\del_q\log\Pi(\tau,\Omega(g))\big\}
\quad\quad\quad\quad &\gcon a\cr
\Pi(\tau,M)=\prod_{n=1}^\infty\det(1-q^nM) &\gcon b\cr}
$$
which follows from \LDh{b}.
The $\Pi$-function in \gcon{b}, which satisfies
$\Pi(\tau,A{\oplus}B)=\Pi(\tau,A)\Pi(\tau,B)$, was first studied by Fegan
\ref\F{H. Fegan, Trans. Am. Math. Soc. {\bf 246} (1978) 339;
J. Diff. Geom. {\bf 13} (1978) 589.}.

For use below, we list a number of simple properties of the
affine-Sugawara characters.
\meti{A.}Evolution operator of $g$. It is convenient to define the
(invertible) evolution operator of $g$,
\eqn\EO{
\Omega_g(\tau,\tau_0,h)=T\e{\int_{\tau_0}^\tau d\tau^\prime
D_g(\tau^\prime,h)}
}
where $T$ denotes $\tau$-ordered product.
This operator is the unique solution of the heat equation and boundary
condition,
\eqna\EOHE
$$
\displaylinesno{
\del\Omega_g(\tau,\tau_0,h)=D_g(\tau,h)\Omega_g(\tau,\tau_0,h)\quad ,\quad
\del_{\tau_0}\Omega_g(\tau,\tau_0,h)=-\Omega_g(\tau,\tau_0,h)D_g(\tau_0,h)
\quad\quad &\EOHE a\cr
\Omega_g(\tau,\tau,h)=1\quad ,\quad\Omega_g(\tau,\tau_0,h)=\Omega_g^{-1}
(\tau_0,\tau,h)\quad\quad &\EOHE b\cr}
$$
and hence $\Omega_g$ determines the evolution of the affine-Sugawara
characters,
\eqn\CG{
\chi_g(T,\tau,h)=\Omega_g(\tau,\tau_0,h)\chi_g(T,\tau_0,h)\ .
}
\meti{B.}Conserved quantities. In addition to the heat equation \heatg{},
the affine-Sugawara characters also satisfy a number of $h$-differential
equations, whose generic form is
\eqna\sing
$$
\displaylinesno{
C^g(T,\tau,h)\chi_g(T,\tau,h)=0&\sing a\cr
C^g(T,\tau',h)=\Omega_g(\tau',\tau,h)C^g(T,\tau,h)\Omega_g(\tau,\tau',h)
\quad .
&\sing b\cr}
$$
The $h$-differential operators $C^g$ are the conserved quantities of the
affine-Sugawara characters.
The simplest example of such relations is the $T$-$\ $and $\tau$-independent
$h$-global Ward identity
\eqn\globid{
Q^g_A(h)\chi_g(T,\tau,h)=0\quad ,\quad Q^g_A(h)=\=E_A(h)+E_A(h)\quad ,\quad
A=1,\ldots ,\dim h
}
but there are other ``spatial" equations \OogI\ of the form \sing{}\ which
follow from the existence of singular vectors in the affine Verma module $V_T$.
The global Ward identity \globid\ tells us that, if desired, we may replace
$E_A\to -\=E_A$ in $D_g$ and more generally on the right side of any term
in $D_{qp}$.
\meti{C.}Explicit form. The explicit form of the affine-Sugawara characters
for integrable representation $T$ of simple $g$ is \Bernard,
\eqna\gsol
$$
\displaylinesno{
\chi_g(T,\tau,h)={1\over \Pi(\tau,\rho(h))\Pi(\tau,\sigma(h))}
\sum_{T'}N_{\T'}^{\T}Tr(h(T'))q^{\Delta_g(T')-{c_g\over 24}}
&\gsol {}\cr}
$$
where the $\Pi$-function is defined in eq. \gcon{b}, and the sum is over
the set of all unitary irreps $T$ of $g$. The coefficients in the sum satisfy
\eqna\NTTP
$$
\displaylinesno{
N_{\T^\prime}^{\T}=\cases{
           \det\omega&if $\lambda(T')=\omega(\lambda(T)+\rho)-\rho+
                                       (x+\~h_g)\sigma $           \cr
               0    &  otherwise\cr}                      &\NTTP {}\cr
}
$$
where $\lambda(T)$ is the highest weight of irrep T, $\omega$ is some element
in the Weyl group of $g$, $\sigma$ is some element
of the coroot lattice, $\rho$ is the Weyl vector, $x$ is the invariant
level and $\~h_g$ is the dual Coxeter number. For $g=\oplus_Ig_I$
and $T=\oplus_IT^I$, the
affine-Sugawara characters are $\chi_g(T)=\prod_I\chi_{g_I}(T^I)$.

\subsec{General Properties of the Connection Moments}

The following properties of the connection moments $D_{qp}$ are easily
established from their definition in \hmoments.
\meti{A.}Representation independence. Since the computational scheme in
subsection 4.2 is independent of irrep $T$ of $g$, the connection moments
$D_{qp}(\~\tau,\tau,h)$ are independent of irrep T.
This means that the representation dependence of the
bicharacters is determined entirely by their affine-Sugawara boundary condition
$\chi(T,\tau,\tau,h)=\chi_g(T,\tau,h)$.
\meti{B.}$\~L$ and $L$ dependence. The one-sided connection moments,
\eqn\Dqz{
D_{q0}(\~L)\quad ,\quad D_{0p}(L)
}
are functions only of $\~L$ and $L$ as shown,
while the mixed moments $D_{qp}(\~L,L)$ with $q,p\ge 1$ are functions of
both $\~L$ and $L$.
\meti{C.}K-conjugation covariance. Under exchange of the K-conjugate ICFT's,
the connection moments exhibit the K-conjugation covariance,
\eqna\Kconjmom
$$
\displaylinesno{
D_{qp}(\~L,L)=D_{pq}(L,\~L)&\Kconjmom a\cr
D_{q0}(\~L)=D_{0q}(L)|_{L\to\~L}\quad .&\Kconjmom b\cr}
$$
\meti{D.}Consistency Relations.
Define the g-covariant derivatives,
\eqn\dG{
d_gf\equiv\del f+fD_g(\tau)\quad ,\quad \~d_gf\equiv\~\del f+fD_g(\~\tau)
}
on any $f(\~\tau,\tau,h)$. Then the connection moments satisfy the consistency
relations,
\eqn\CREL{
d_gD_{qp}=D_{q+1,p}+D_{q,p+1}\quad ,\quad D_{00}=1
}
in analogy to the consistency relations on the sphere \obersI.
When $q=p=0$, these relations reproduce the identity
$D_g=D_{10}+D_{01}$ in eq. \heatg{}.
Following the development on the sphere, the consistency relations
can be solved at each fixed value of $q+p$ to express all $D_{qp}$ in terms
of the canonical sets $\{D_g,D_{0p}\}{\rm\ or\ }\{D_g,D_{q0}\}$ .
\meti{E.}Other relations. It appears that all the relations known for
the connection moments on the sphere \refs{\obersI,\obersII} have their close
counterparts on the torus. Among these, we list only the translation sum rule,
\eqn\transumrule{
\sum_{r,s=0}^\infty {(\tau-\tau_0)^{r+s}\over r!s!}D_{r+q,s+p}(\tau_0)=
D_{qp}(\tau)\Omega_g(\tau,\tau_0,h)
}
where $\Omega_g$ is the evolution operator of $g$, and the
partially-factorized form of the bicharacters,
\eqna\partfact
$$
\displaylinesno{
\chi(T,\~\tau,\tau,h)=\sum_{q,p=0}^\infty {(\~\tau-\tau_0)^{q}\over q!}
C_{qp}(T,\tau_0,h) {(\tau-\tau_0)^{p}\over p!}&\partfact a\cr
C_{qp}(T,\tau_0,h)\equiv D_{qp}(\tau_0,h)\chi_g(T,\tau_0,h)&\partfact b\cr}
$$
which follows from the Ward identities using \transumrule. The right side of
\partfact{a}\ is independent of the regular reference point $\tau_0$.

Following the development on the sphere \obersI, the eigenvectors of the matrix
$C_{qp}(\tau_0)$ give a factorization of the bicharacters,
\eqn\fact{
\chi(T,\~\tau,\tau,h)=\sum_\nu\chi_{\~L}^\nu(T,\~\tau,\tau_0,h)
\chi_L^\nu(T,\tau,\tau_0,h)
}
into candidate conformal characters $\chi_{\~L}$ and $\chi_L$ of the
$\~L$ and the $L$ theory respectively. On the sphere, the corresponding
conformal correlators of $\~L$ and $L$ are covariant under the braid group, so
the analogous form \fact\ should be studied for modular covariance on the
torus.
We will return to the subject of factorization in Sections 7 and 9.

\newsec{Flat Connections on the Torus}

Following Ref. \obersIII, we may reexpress the Ward identities \AVWI{}\
as heat-like differential equations with flat connections.

Using Taylor series in $\~\tau$ or $\tau$ and the Ward identities, we first
write the bicharacters in the  two equivalent forms,
\eqna\WIT
$$
\displaylinesno{
\vphantom{\Biggl(}
\chi(T,\~\tau,\tau,h)=\~{B}(\~\tau,\tau,h)\chi_g(T,\tau,h)=
{B}(\~\tau,\tau,h)\chi_g(T,\~\tau,h)\quad &\WIT a\cr
\~{B}(\~\tau,\tau,h)=
\sum_{q=0}^\infty{(\~\tau-\tau)^q\over q!}D_{q0}(\tau,h)\quad ,\quad
{B}(\~\tau,\tau,h)=
\sum_{p=0}^\infty{(\tau-\~\tau)^p\over p!}D_{0p}(\~\tau,h)\ .\quad &\WIT b\cr}
$$
These forms show explicitly  that the affine-Virasoro characters $\chi$
are completely determined given the affine-Sugawara characters $\chi_g$ and
the connection moments, which appear in the (invertible) $h$-differential
operators
$\~{B}$ and $B$.

Then, by differentiation of \WIT{}, we obtain the heat-like differential
equations for the bicharacters,
\eqna\HE
$$
\displaylinesno{
\~\del\chi(T,\~\tau,\tau,h)=\~D(\~\tau,\tau,h)\chi(T,\~\tau,\tau,h)&\HE a\cr
\del\chi(T,\~\tau,\tau,h)=D(\~\tau,\tau,h)\chi(T,\~\tau,\tau,h)&\HE b\cr
\~D=\~\del\~{B}{\~{B}}^{-1} \quad ,\quad
D=\del{B}{B}^{-1}&\HE c\cr}
$$
where the $h$-differential operators $\~D$ and $D$ are the affine-Virasoro
connections
of the heat-like system. Eq. \HE{c}\ defines the connections as non-linear
functionals of the connection moments, and the connections may also be
evaluated in principle from the formulae,
\eqna\alternativeD
$$
\eqalignno{
\~D(\~\tau,\tau,h)\chi(T,\~\tau,\tau,h)&=
2\pi i\Tr_{\T}\left(\~q^{\~L(0)-\~c/24}q^{L(0)-c/24}
\big(\~L(0)-\~c/24\big)h\right)&\alternativeD a\cr
D(\~\tau,\tau,h)\chi(T,\~\tau,\tau,h)&=
2\pi i\Tr_{\T}\left(\~q^{\~L(0)-\~c/24}q^{L(0)-c/24}
\big(L(0)-c/24\big)h\right)&\alternativeD b\cr}
$$
which follow from \bichar\ and the heat-like system \HE{}.

We note that, like the connection moments, the operators
$\~{B}$, ${B}$ and the connections $\~D$, $D$ are
independent of the representation $T$.

Following the development of Ref. \obersIII, we find close analogues of
the general relations known for the connections on the sphere.
\meti{A.}Flat connections. To see that the connections $\~D,D$ are flat,
define covariant derivatives,
\eqn\FID{
\~df \equiv\~\del f+f\~D\quad ,\quad df\equiv\del f+fD\quad ,\quad\forall f
\quad .}
Then the flatness condition,
\eqn\sndcrossflat{
d\~D=\~d D
}
is obtained by differentiation of the heat-like equations.
The same condition is also obtained by differentiation of \alternativeD{},
using the heat-like equations.
\meti{B.}Inversion formula. Note that the covariant derivatives
commute
\eqn\FCI{
[d,\~d]=0
}
because the connections are flat,
and that the inversion formula for the connection moments,
\eqn\didibar{
D_{qp}(\tau,h)={\~d}^qd^p1|_{\~\tau=\tau}
}
follows by multiple differentiation of the heat-like system. The
formula \didibar, which is the inverse of \HE{c}, expresses the
connection moments in terms of the flat connections. As examples, we
use the formula to list the first few moments,
\eqna\momex
$$
\displaylinesno{
D_{00}(\tau)=1&\momex a\cr
D_{10}(\tau)=\~D(\tau,\tau)\quad ,\quad D_{01}(\tau)=D(\tau,\tau)&\momex b\cr
D_{20}(\tau)=(\~\del\~D+{\~D}^2)|_{\~\tau=\tau} \quad ,\quad
D_{02}(\tau)=(\del D+D^2)|_{\~\tau=\tau} &\momex c\cr
D_{11}(\tau)=(\~\del D+D\~D)|_{\~\tau=\tau}=(\del\~D+\~D D)|_{\~\tau=\tau}.
&\momex d\cr}
$$
As on the sphere, we note that the pinched connections
(at $\~\tau=\tau$) are always equal to the first connection moments.
\meti{C.}Evolution operators.  It follows from \WIT{}\ and \HE{c}\ that
the operators  $\~{B}$, ${B}$ in
\WIT{}\ are the (invertible) evolution operators of the flat connections,
\eqna\caldevolution
$$
\eqalignno{
\~{B}(\~\tau,\tau,h)&=\~T\e{\int_{\tau}^{\~\tau}d\~\tau'
\~D(\~\tau',\tau,h)}&\caldevolution a\cr
{B}(\~\tau,\tau,h)&=T\e{\int_{\~\tau}^\tau d\tau' D(\~\tau,\tau',h)}
&\caldevolution b\cr
\~\del\~{B}=\~D\~{B} &\quad ,\quad
\del{B}=D{B} &\caldevolution c\cr
\~{B}(\tau,\tau,h)&={B}(\~\tau,\~\tau,h)=1\quad .&\caldevolution d\cr}
$$
where $\~T$ and $T$ are $\~\tau$ and $\tau$ ordering respectively.
Moreover, the two forms of the bicharacter in \WIT{a}\ show that the
evolution operators of the flat connections are related by the
evolution operator of $g$,
\eqn\caldprop{
\~{B}(\~\tau,\tau,h)=
{B}(\~\tau,\tau,h)
\Omega_g(\~\tau,\tau,h)\quad ,\quad
{B}(\~\tau,\tau,h)=
\~{B}(\~\tau,\tau,h)
\Omega_g(\tau,\~\tau,h)
}
and hence the evolution operator of $g$ is composed of the evolution
operators of the flat connections,
\eqn\Omfact{
\Omega_g(\~\tau,\tau,h)=
{B}^{-1}(\~\tau,\tau,h)\~{B}(\~\tau,\tau,h)\quad .
}
The identities \caldprop\ also imply the differential relations
\eqn\newrel{
(\~d_g-\~D){B}=(d_g-D)\~{B}=0
}
which supplement the differential relations in \caldevolution{c}.
\meti{D.}$\~L$ and $L$ dependence. According to \WIT{b}\ and \Kconjmom{}, the
evolution operators $\~{B}(\~L)$ and ${B}(L)$ and the
connections $\~D(\~L)$ and $D(L)$  are functions of only $\~L$ and $L$
as shown.
\meti{E.}K-conjugation covariance. The evolution operators and connections
satisfy the K-conjugation covariance,
\eqn\Kconjcov{
{B}(L,\~\tau,\tau,h)|_{\tau\lra\~\tau\atop L\to\~L}=
\~{B}(\~L,\~\tau,\tau,h)\quad ,\quad
D(L,\~\tau,\tau,h)|_{\tau\lra\~\tau\atop L\to\~L}=\~D(\~L,\~\tau,\tau,h)\quad .
}
\meti{F.}Non-local conserved quantities \obersIII. Following subsection 4.3 and
the development on
the sphere, we find a non-local conserved quantity $C(T,\~\tau,\tau,h)$,
\eqna\nonlocal
$$
\eqalignno{
&C(T,\~\tau,\tau,h)\chi(T,\~\tau,\tau,h)=0 &\nonlocal a\cr
&C={B}C_g(\~\tau){B}^{-1}
=\~{B}C_g(\tau)\~{B}^{-1} &\nonlocal b\cr}
$$
for each of the conserved quantities $C_g(T,\tau,h)$ of the affine-Sugawara
character on an $h$ source. These non-local conserved quantities are the
lift of the  conserved quantities $C_g$ into the space of ICFT's.

Because they are related by a similarity
transformation, the algebra of the non-local quantities is the same as the
algebra of the $C_g$'s. For example, we have
the non-local conserved generators $Q_A(\~\tau,\tau,h)$ of $h\subset g$,
\eqna\simtrans
$$
\displaylinesno{
Q_A={B}Q_A^g{B}^{-1}=
\~{B}Q_A^g\~{B}^{-1}&\simtrans a\cr
Q^{\phantom g}_A\chi=0\quad ,\quad [Q_A,Q_B]=if_{AB}^{\ \ \ C}Q_C
&\simtrans b\cr}
$$
where $Q_A^g=\=E_A+E_A$, $A=1,\ldots ,\dim h$ are the global generators
of $h\subset g$. As on the sphere, the non-local generators of $h\subset g$
degenerate to the global generators of $h\subset g$ when $h$ is an ordinary
Lie symmetry of the K-conjugate pair. We will check this explicitly for $h$
and the $g/h$ coset constructions in the following section, but it is also true
for all the Lie $h$-invariant CFT's \Lieh.

We also find non-local conserved quantities $C$ associated to the $C_g$'s of
the null states of the affine modules \OogI. These conserved quantities provide
further differential
relations on the bicharacters, beyond the heat equations, but we will not
need their explicit form here. Instead, we encode  their information in the
bicharacters by the choice of the correct affine-Sugawara characters $\chi_g$
in the affine-Sugawara boundary condition.

\newsec{Coset Constructions}

\subsec{Choosing the Source}

In this Section, we study the bicharacters of $h$ and the $g/h$ coset
constructions \refs{\BH,\H,\GKO},
\eqn\neweq{
\~L=L_{g/h}\quad ,\quad L=L_h
}
with $G/H$ a reductive coset space. In this case,
we are able to solve the system exactly  by choosing the subgroup $H$
of the source $h$ to be the {\it same} subgroup involved in the $G/H$ coset.
Then, the bicharacter and its heat-like equations have the form,
\eqna\cosetbchi
$$
\displaylinesno{
\chi(T,\~\tau,\tau,h)=Tr_{\T}\left(\~q^{L_{g/h}(0)-c_{g/h}/24}
q^{L_h(0)-c_h/24}h\right)&\cosetbchi a\cr
\~\del\chi=\~D(L_{g/h})\chi\quad ,\quad\del\chi=D(L_h)\chi&\cosetbchi b\cr}
$$
where $\~D(L_{g/h})$ and $D(L_h)$ are the connections of $g/h$ and
$h$ respectively.
Evaluation of the bicharacters of $h$ and $g/h$ on sources
larger than $H$ is more complicated, as discussed below and in the following
Section.

\subsec{The Subgroup Connection}

We evaluate $D(L_h)$ from the general form \alternativeD{b}, which now reads,
\eqna\dlh
$$
\displaylinesno{
D(L_h,\tau,h)\chi(T,\~\tau,\tau,h)=
2\pi i Tr_{\T}\left(\~q^{L_{g/h}(0)-c_{g/h}/24}
q^{L_h(0)-c_h/24}(L_h(0)-{c_h\over 24})h\right)\vphantom{\Bigl(}\quad &\dlh
a\cr
L_h(0)=L^{AB}_h\big(J_A(0)J_B(0)+2\sum_{n>0}J_A(-n)J_B(n)\big)
\vphantom{\vrule height 15pt depth 5pt width 0pt}
\quad ,\quad L_{g/h}(0)=L_g(0)-L_h(0)\quad &\dlh b\cr}
$$
where $A,B=1,\ldots ,\dim h$ and $L_g(0)$ is given in eq. \Lg.

In this case, we are able to follow the strategy of subsection 4.2 exactly.
We need the relations,
\eqna\relII
$$
\eqalignno{
q^{L_h(0)}J_A(-n)&=q^nJ_A(-n)q^{L_h(0)}&\relII a\cr
\~q^{L_{g/h}(0)}J_A(-n)&=J_A(-n)\~q^{L_{g/h}(0)}&\relII b\cr
hJ_A(-n)&=\rho(h)_A^{\ \ B}J_B(-n)h&\relII c\cr
E_A h&=hJ_A(0)&\relII d\cr}
$$
where \relII{a,b}\ records that
the $h$-currents $J_A$ have conformal weights 1 and 0 respectively
under $T_h$ and $T_{g/h}$. Then, we obtain the formulae,
\eqna\comtrick
$$
\displaylinesno{
Tr_{\T}\left(\~q^{L_{g/h}(0)}q^{L_h(0)}J_A(-n){\cal O}h\right)\hfill & \cr
\hfill {}=\left({q^n\rho(h)\over 1-q^n\rho(h)}\right)_A^{\ B}
Tr_{\T}\left(\~q^{L_{g/h}(0)}q^{L_h(0)}
\big[{\cal O},J_B(-n)\big]h\right)\ ,\ n\ne 0\quad\quad\quad\quad&\comtrick
a\cr
\vphantom{\Biggl(}
Tr_{\T}\left(\~q^{L_{g/h}(0)}q^{L_h(0)}J_A(0){\cal O}h\right)=E_A
Tr_{\T}\left(\~q^{L_{g/h}(0)}q^{L_h(0)}{\cal O}h\right)\hfill &\comtrick b\cr}
$$
for any operator ${\cal O}$. Note that, had we chosen a source in $G$ or
a non-reductive coset space, eq. \relII{c}\ would have contained extra terms
with coset currents $J_I$, so that averages of the $h$-currents no longer
satisfy closed equations. We will return to the subject of sources larger then
$H$ in Section 9.

Using \dlh{}\ and \comtrick{}, we obtain the exact $h$-connection,
\eqna\DLH
$$
\displaylinesno{
\qquad D(L_h,\tau,h)= 2\pi i\big\{ -c_h/24+
L_h^{AB}E_A(h)E_B(h)\hfill & \cr
\hfill {}+2L_h^{AB}
\sum_{n>0}\big({q^n\rho(h)\over 1-q^n\rho(h)}\big)_A^{\ C}
(if_{BC}^{\ \ D}E_D(h)+nG_{BC})\big\}\ .\quad\quad\quad\quad &\DLH {}\cr}
$$
The $h$-connection is not a function of
$\~\tau$, so that, according to eq. \momex{b}, the $h$-connection is
equal to the first connection moment of the $h$-theory,
\eqn\DH{
D_h(\tau,h)\equiv D(L_h,\tau,h)=D_{01}(L_h,\tau,h)\quad .
}
This relation is easily checked by comparison with the general
first moment result \DOlII{}\ in this case.

We also observe the embedding relation between the connections of $g$ and $h$,
\eqn\embed{
D_h(L_h,\tau,h)=D_g(L_g,\tau,g)|_{L_g\to L_h\atop g\to h}
}
which follows on comparison of eqs. \heatBER{b}\ and \DLH{}.
Following Ref. \obersIII,
such embedding relations can be used to compute the connections of all the
affine-Sugawara nests on $g\supset h_1\dots\supset h_n$, but we will limit
our work here to the simplest case
of the coset constructions.

\subsec{ The Coset Connection}
Having determined the $h$-connection $D(L_h)=D_h$ by direct
computation, we may obtain the coset connection $\~D(L_{g/h})$ by solving
the flatness condition \sndcrossflat, which now reads
\eqn\crossDh{
\del\~D(L_{g/h})=[D_h,\~D(L_{g/h})]
}
because $\~\del D_h=0$.
The flatness condition \crossDh\ determines the $\tau$ dependence of
the coset connection as
\eqn\Dbar{
\~D(L_{g/h},\~\tau,\tau,h)=
\Omega_h(\tau,\~\tau,h)\~D(L_{g/h},\~\tau,\~\tau,h)
\Omega_h^{-1}(\tau,\~\tau,h)
}
where
$\Omega_h$ is the (invertible) evolution operator of $h$,
which satisfies,
\eqna\omegh
$$
\displaylinesno{
\Omega_h(\tau,\~\tau,h)=T\e{\int_{\~\tau}^\tau d\tau' D_h(\tau',h)}
&\omegh a\cr
\del\Omega_h(\tau,\~\tau,h)=D_h(\tau,h)\Omega_h(\tau,\~\tau,h)\quad ,\quad
\~\del \Omega_h(\tau,\~\tau,h)= -\Omega_h(\tau,\~\tau,h)D_h(\~\tau,h)
\quad &\omegh b\cr
\Omega_h(\~\tau,\~\tau,h)=1\quad ,\quad
\Omega_h^{-1}(\tau,\~\tau,h)=\Omega_h(\~\tau,\tau,h)\quad
&\omegh c\cr}
$$
in analogy to the evolution operator of $g$.
The quantity $\~D(L_{g/h},\~\tau,\~\tau,h)$ in \Dbar\ is the pinched
coset connection, whose form is known from eqs. \momex{b}\ and \dlh{b},
\eqn\Dbarpinch{
\~D(L_{g/h},\~\tau,\~\tau,h)=D_{10}(L_{g/h},\~\tau,h)
= D_g(\~\tau,h)-D_h(\~\tau,h)\equiv D_{g/h}(\~\tau,h)\quad .
}
Combining eqs. \Dbar\ and \Dbarpinch, we obtain the coset connection,
\eqn\Dbarsol{
\~D(L_{g/h},\~\tau,\tau,h)=\Omega_h(\tau,\~\tau,h)D_{g/h}(\~\tau,h)
\Omega_h^{-1}(\tau,\~\tau,h)
}
which, as on the sphere \refs{\obersI{--}\obersIII}, is an $h$-dressing of
the first coset connection moment.

\subsec{Connection Moments of $h$ and $g/h$}

Having obtained the $h$ and $g/h$ connections $D(L_h)=D_h$ and
$D(L_{g/h})$, we may compute the connection moments $D_{qp}$ of $h$
and $g/h$ from the inversion formula \didibar,
\eqn\invform{
D_{qp}={\~d}^qd^p1|_{\~\tau=\tau}\quad .
}
In this case,  the inversion formula simplifies to the factorized form,
\eqna\dqpfact
$$
\displaylinesno{
D_{qp}(\tau)=D_{0p}^h(\tau)D_{q0}^{g/h}(\tau)&\dqpfact a\cr
D_{0p}^h(\tau)\equiv d^p1|_{\~\tau=\tau}\quad ,\quad
D_{q0}^{g/h}(\tau)\equiv {\~d}^q1|_{\~\tau=\tau}&\dqpfact b\cr}
$$
because $\~\del D_h=0$. The result \dqpfact{}\ is analogous to the
factorization of the connection moments of $h$ and $g/h$ on the sphere
\refs{\obersI,\obersII}.

Computation of the moments from the factorized form \dqpfact{}\ is
particularly simple, and we list the examples,
\eqna\sndmoment
$$
\displaylinesno{
D_{01}^h=D_h\quad ,\quad
D_{10}^{g/h}= D_{g/h}&\sndmoment a\cr
D_{02}^h=\del D_h+D_h^2\quad ,\quad
D_{11}= D_hD_{g/h}&\sndmoment b\cr
D_{20}^{g/h}=\del D_{g/h}+D_{g/h}^2 +[D_{g/h},D_h]&\sndmoment c\cr}
$$
through order $q+p\le 2$.

\subsec{The Bicharacters of $h$ and $g/h$}

Given the flat connections $D(L_h)$ and $D(L_{g/h})$, the bicharacters
$\chi(\~\tau,\tau)$ are the unique solution to the heat-like
system \cosetbchi{b}\ with the affine-Sugawara boundary condition
$\chi(\tau,\tau)=\chi_g(\tau)$.
To find this solution quickly, use $D=D_h$ and eq. \caldprop\ to obtain
the evolution operators of the flat connections,
\eqn\caldeqOM{
{B}(\~\tau,\tau,h)=\Omega_h(\tau,\~\tau,h)\quad ,\quad
\~{B}(\~\tau,\tau,h)=\Omega_h(\tau,\~\tau,h)\Omega_g(\~\tau,\tau,h)
\quad .
}
Then the bicharacters of $h$ and $g/h$,
\eqn\Dcalomeg{
\chi(T,\~\tau,\tau,h)=
\Omega_h(\tau,\~\tau,h)\chi_g(T,\~\tau,h)
}
follow immediately as a special case of the general result \WIT{a}.

Using the heat equations \heatg{}\ and \omegh{}\ for $\Omega_h$ and $\chi_g$,
it is easy to check directly that the bicharacters \Dcalomeg\ solve the
heat-like system \cosetbchi{b}. We find,
\eqna\delbchi
$$
\eqalignno{
\del\chi&=D(L_h)\chi &\delbchi a\cr
\~\del\chi&=\Omega_h(\tau,\~\tau,h)D_{g/h}(\~\tau,h)\chi_g(T,\~\tau,h)
&\delbchi b\cr
&=\Omega_h(\tau,\~\tau,h)D_{g/h}(\~\tau,h)
\Omega_h^{-1}(\tau,\~\tau,h)\chi(T,\~\tau,\tau,h)&\delbchi c\cr
&=D(L_{g/h})\chi &\delbchi d\cr}
$$
where the flat connections $D_h=D(L_h)$ and $\~D(L_{g/h})$ are given in
eqs. \DLH{}\ and \Dbarsol.

\subsec{Non-local Conserved Quantities}

Using the $h$-transformation properties of the adjoint representation,
\eqn\Erho{
\=E_A\rho(h)_B^{\ C}=\rho(h)_B^{\ D}(T_A^{adj})_D^{\ C}\quad ,\quad
E_A\rho(h)_B^{\ C}=-(T_A^{adj})_B^{\ D}\rho(h)_D^{\ C}
}
we verify the $h$-invariance of the $h$-evolution operator and the
connections of $h$ and $g/h$,
\eqna\conhinv
$$
\displaylinesno{
[Q_A^g(h),D_h(\tau,h)]=
[Q_A^g(h),\Omega_h(\tau,\~\tau)]=
[Q_A^g(h),\~D(L_{g/h},\~\tau,\tau)]=0\quad &\conhinv {}\cr}
$$
where $Q^g_A(h)=\=E_A(h)+E_A(h)$ are the global generators of $h\subset g$.
Then, the non-local conserved generators \simtrans{}\ of $h$ and $g/h$,
\eqna\nonlCA
$$
\displaylinesno{
Q_A(\~\tau,\tau,h)\chi(T,\~\tau,\tau,h)=0\quad ,\quad
A=1,\ldots ,\dim h&\nonlCA a\cr
Q_A(\~\tau,\tau,h)=
\Omega_h(\tau,\~\tau,h)Q^g_A(h)\Omega_h^{-1}(\tau,\~\tau,h)=Q^g_A(h)
&\nonlCA b\cr}
$$
are equal to the global generators of $h\subset g$. If we choose the source
in $G$, we still find $Q^{\phantom g}_A=Q^g_A$ because $\~{B}$ and
${B}$ are $h$-invariant.  On the other hand,
the extra conserved coset generators $Q_I={B}Q^g_I{B}^{-1},
\ \ I=1,\ldots ,\dim g/h$ remain non-local on the $G$ source, in parallel
with results on the sphere \obersIII.

\newsec{Integral Representation for Coset Characters}

To further analyze the bicharacters of $h$ and $g/h$, we introduce the
$\^h$-characters for integrable representation $T^h$ on an $h$ source,
\eqna\hathchar
$$
\eqalignno{
\chi_h(T^h,\tau,h)&=Tr_{\T^h}\big(q^{L_h(0)-c_h/24}h\big)&\hathchar a\cr
&= {1 \over \Pi (\tau,\rho(h))}
\sum_{T'^h} N_{\T'^h}^{\T^h}Tr(h(T'^h))
q^{\Delta_h(T'^h)-c_h/24} &\hathchar b\cr}
$$
where $h$ is a simple subalgebra of $g$, the sum is over all the unitary irreps
of $h$ and $N_{\T'^h}^{\T^h}$ is the $h$-analogue of $N_{\T'}^\T$ in \NTTP{}.
The connection between the characters of $T$ (irrep of $g$) and $T^h$
(irrep of $h$) is
\eqn\multip{
Tr(h(T))=\sum_{T^h}m(T,T^h)Tr(h(T^h))
}
where $m(T,T^h)$ is the multiplicity of irrep $T^h$ in irrep $T$.

The $\^h$-characters satisfy the heat and evolution equations,
\eqna\Hheat
$$
\eqalignno{
\del\chi_h(T^h,\tau,h)&=D_h(\tau,h)\chi_h(T^h,\tau,h)&\Hheat a\cr
\chi_h(T^h,\tau',h)&=\Omega_h(\tau',\tau,h)\chi_h(T^h,\tau,h)&\Hheat b\cr}
$$
where $D_h$ is the same $h$-connection which controls the $\tau$ dependence
of the bicharacter.
It is therefore reasonable to assume that the bicharacter lives in the space
spanned by the $\^h$-characters,
\eqn\hexpan{
\chi(T,\~\tau,\tau,h)=\sum_{T^h}\vphantom{\Biggl(}^\prime
\chi_{g/h}(T,T^h,\~\tau)\chi_h(T^h,\tau,h)
}
which solves the bicharacter equation $\del\chi=D_h\chi$
so long as the coset characters $\chi_{g/h}$ are independent of the source.
In \hexpan, the primed sum is over the integrable representations $T^h$ of $h$
at the induced level of the subalgebra.

\nref\KW{V.G. Ka\v c and M. Wakimoto, Adv. Math. {\bf 70} (1988) 156.}
\nref\GKOII{P. Goddard, A. Kent and D. Olive, Commun. Math. Phys. {\bf 103}
(1986) 105.}
\nref\Gepner{D. Gepner and Z. Qiu, Nucl. Phys. {\bf B285 [FS19]} (1987) 423.}
At $\tau{=}\~\tau$, the factorized form \hexpan\ implies the known
factorization of the affine-Sugawara characters
\refs{\KW,\GKOII,\Gepner},
\eqn\KW{
\chi_g(T,\~\tau,h)=\sum_{T^h}\vphantom{\Biggl(}'
\chi_{g/h}(T,T^h,\~\tau)\chi_h(T^h,\~\tau,h)
}
and, conversely, using \KW\ in the bicharacter solution \Dcalomeg, we recover
\eqna\rederiv
\hexpan\ in the steps,
$$
\eqalignno{
\chi(T,\~\tau,\tau,h)&=\Omega_h(\tau,\~\tau,h)
\sum_{T^h}\vphantom{\Bigl(}'
\chi_{g/h}(T,T^h,\~\tau)\chi_h(T^h,\~\tau,h)&\rederiv a\cr
&=\sum_{T^h}\vphantom{\Bigl(}'
\chi_{g/h}(T,T^h,\~\tau)\Omega_h(\tau,\~\tau,h)
\chi_h(T^h,\~\tau,h)&\rederiv b\cr
&=\sum_{T^h}\vphantom{\Bigl(}'
\chi_{g/h}(T,T^h,\~\tau)\chi_h(T^h,\tau,h)\quad .&\rederiv c\cr}
$$
To obtain \rederiv{b}, we used the fact that the coset characters are
independent of the source.

In order to check that the factorized form \hexpan\ also satisfies the
$\~D$ equation, follow the steps,
\eqna\steps
$$
\eqalignno{
\Omega_h(\~\tau,\tau,h)&\big(\~\del-\~D(L_{g/h},\~\tau,\tau,h)\big)
\chi(T,\~\tau,\tau,h) & \steps a\cr
&=\sum_{T^h}\vphantom{\Bigl(}'
\big(\~\del\chi_{g/h}(T,T^h,\~\tau)-\chi_{g/h}(T,T^h,\~\tau)
D_{g/h}(\~\tau,h)\big)\chi_h(T^h,\~\tau,h)\quad\quad  &\steps b\cr
&=\big(\~\del-D_g(T,\~\tau,h)\big)\chi_g(T,\~\tau,h)=0 &\steps c\cr}
$$
where we used the form \Dbarsol\ of the coset connection $\~D(L_{g/h})$
and the heat
equations on $g$ and $h$. The equation $\~\del\chi=\~D\chi$ is then
satisfied because $\Omega_h$ is invertible.

We may now obtain linear differential equations for the coset characters
as follows. From the definition \NTTP{}\ of the coefficients
$N_{\T'^h}^{\T^{h}}$, it follows that \ref\GK{K. Gaw\c edzki and A. Kupiainen,
Nucl. Phys. {\bf B320} (1989) 625},
\eqn\NNeqDEL{
N_{\T''^{h}}^{\T^h} N_{\T''^{h}}^{\T'^{h}}=\delta(T^h,T'^h)
|N_{\T''^{h}}^{\T^h}|
}
where $\delta$ is Kronecker delta,
and we know that
\eqn\charortho{
\int dh\ Tr\big(h^*(T'^{h})\big)Tr\big(h(T^{h})\big)=\delta(T'^h,T^{h})
}
where $dh$ is Haar measure on $h$. Using \NNeqDEL\ and \charortho,
we verify the orthonormality relation for $\^h$-characters,
\eqna\afcharortho
$$
\displaylinesno{
\int dh \chi_h^\dagger (T'^{h},\tau,h)\chi_h(T^{h},\tau,h)
=\delta(T'^h,T^h)&\afcharortho a\cr
\chi_h^\dagger(T^h,\tau,h)\equiv
{\Pi(\tau,\rho(h))\over f(T^h,q)}
\sum_{T'^h} N_{T'^h}^{T^h}Tr(h^*(T'^h))
q^{\Delta_h(T'^h)+c_h/24}&\afcharortho b\cr
f(T^h,\tau)\equiv\sum_{T'^h} |N_{T'^h}^{T^h}|
q^{2\Delta_h(T'^h)}\quad .&\afcharortho c\cr}
$$
With $h\to g$, $\rho(h)\to\Omega(g)$ and $T^h\to T$, these relations apply
as well for the $\^g$-character $\chi_g(T,\tau,g)$ on a G source.

Then, integrating eq. \steps{b}\ with  $\int
dh\chi_h^{\dagger}(T'^h,\~\tau,h)$,
we obtain the coset equations,
\eqna\newcoseteq
$$
\displaylinesno{
\~\del\chi_{g/h}(T,T^h,\~\tau)=\sum_{T'^h}^{\phantom T}\vphantom{\Bigl(}'
w[L_{g/h},T^h,T'^h,\~\tau]\chi_{g/h}(T,T'^h,\~\tau) &\newcoseteq a\cr
w[L_{g/h},T^h,T'^h,\~\tau]=
\int dh\chi_h^{\dagger}(T^h,\~\tau,h)D_{g/h}(\~\tau,h)\chi_h(T'^h,\~\tau,h)
&\newcoseteq b\cr}
$$
where $D_{g/h}$, given in eq. \Dbarpinch, is the first connection moment of
the coset construction.  These equations are the analogue of the coset
equations
in the block basis on the sphere \obersI, and we note that, as on the sphere,
the c-function coset coefficients $w[L_{g/h}]$ in \newcoseteq{b}\
are an $h$-dressing of the first coset connection moment.

The correct solutions of the coset equations (which respect the affine cutoff
of $\^g$ and $\^h$) are most easily obtained by the same
projection on eq. \KW.
We obtain an integral representation for the general $g/h$ coset character,
\eqna\cosetchar
$$
\displaylinesno{
\chi_{g/h}(T,T^h,\~\tau)=\int dh\chi_h^\dagger(T^h,\~\tau,h)\chi_g(T,\~\tau,h)
\hfill &\cosetchar a\cr
\quad\ ={\~q^{-{c_{g/h}\over 24}}\over f(T^h,\~\tau) }
\sum_{T', T'^h}
N_{T'}^{T}N_{T'^h}^{T^h}
\left(\int dh{Tr(h^*(T'^h))Tr(h(T'))\over\Pi(\~\tau,\sigma(h))}\right)
\~q^{\Delta_g(T')+\Delta_h(T'^h)}\ . \hfill &\cosetchar b\cr}
$$
The general result \cosetchar{a}, which we have been unable to find in the
literature, holds for semi-simple $g$ and simple $h$. In form, this result is
the analogue of the formula ${\cal C}_{g/h}={\cal F}_g{\cal F}_h^{-1}$ for
the coset blocks on the sphere \refs{\Doug,\obersI,\obersII}. The special case
in \cosetchar{b}\ is the explicit form of \cosetchar{a}\ for simple $g$.

\newsec{High-level Affine-Virasoro Characters}

\subsec{High-level Systematics}

In this Section, we consider the high-level affine-Virasoro characters
for the broad class of ICFT's which are high-level smooth
on simple $g$. In this case, the high-level forms of the inverse inertia
tensors are \refs{\HO,\HYII},
\eqna\highkLab
$$
\displaylinesno{
\~L^{ab}={\~P^{ab}\over 2k}+O(k^{-2})\quad ,\quad
L^{ab}={P^{ab}\over 2k}+O(k^{-2})&\highkLab a\cr
\~c=\rank\~P+O(k^{-1})\quad ,\quad c=\rank P+O(k^{-1}) &\highkLab b\cr}
$$
where $\~P$ and $P$ are the high-level projectors of the $\~L$ and the $L$
theory respectively.

According to \hmoments, \highkLab{}\ and the
affine algebra \comrelg{}, we see that each new factor $\~L$ or $L=O(k^{-1})$
in $D_{qp}$ comes with two more currents and hence with the possibility
of one more current contraction, proportional to the central term
$G_{ab}=k\eta_{ab}$.
It follows that all $D_{qp}$ begin at the same order of $k^{-1}$.
Since $D_{00}=1$, we conclude that $D_{qp}$, $\~D$ and $D$ are power
series in $k^{-1}$ with leading terms,
\eqn\orderk{
\left\{D_{qp},\~D,D\right\}=O(k^0)
}
which come entirely from current contractions.

\subsec{High-level Flat Connections}

More explicitly, we will evaluate the high-level form of the flat connection
$\~D$, using the high-level form of \alternativeD{},
\eqn\highkD{
\eqalign{
&\~D(\~\tau,\tau,g)\chi(T,\~\tau,\tau,g)\bbuildrel{=}_{k}^{}\cr
&2\pi iTr_{\T}\left({\~q}^{\~L(0)-\rank\~P/24}q^{L(0)-\rank P/24}
\big({\~P^{ab}\over k}\sum_{n>0}J_a(-n)J_b(n)-{\rank\~P\over 24}\big)
g\right)\cr}
}
for general source $g\in G$. Here, we have already used the high-level form
of $\~L^{ab}$ and we have neglected the zero-mode
contribution of the currents, which is automatically higher order.

To proceed, we need the high-level form of the $T,J$ commutator in \LJ{},
\eqn\highkcomrel{
[L(0),J_a(-n)]\bk n\big(PJ(-n)\big)_a\quad ,\quad
[\~L(0),J_a(-n)]\bk n\big(\~PJ(-n)\big)_a
}
where $(PJ)_a\equiv P_a^{\ b}J_b$ and $P_a^{\ b}\equiv\eta_{ac}P^{cb}$.
This gives the high-level analogue of eq. \FII{a},
\eqn\highkfirstid{
\~q^{\~L(0)}q^{L(0)}J_a(-n)\bk \big( (\~q^n\~P+q^nP)J(-n)\big)_a
\~q^{\~L(0)}q^{L(0)}
}
and then we may follow the usual procedure to express the $J_aJ_b$  trace
in terms of the commutator of the two currents. Keeping only the  contraction
term in the commutator, we obtain the relation,
\eqn\highkjajb{
Tr_{\T}\left(\~q^{\~L(0)-c/24}q^{L(0)-c/24}J_a(-n)J_b(n)g\right)\bk
kn\left\{\big(
{(\~q^n\~P+q^nP)\Omega(g)\over 1-(\~q^n\~P+q^nP)\Omega(g)}
\big)_a^{\ c}\eta_{cb}
\right\}\chi(T,\~\tau,\tau,g)
}
where $\Omega(g)=g^{-1}(T^{adj})$ is the adjoint representation of $G$.

Using \highkD\ and \highkjajb, we read off the leading terms of the flat
connections,
\eqna\highkflatc
$$
\displaylinesno{
\~D(\~L,\~\tau,\tau,g)=2\pi i\left(\sum_{n>0}nTr\big({X_n\over 1-X_n}\~P\big)
-{\rank\~P\over 24}\right)+O(k^{-1})&\highkflatc a\cr
D(L,\~\tau,\tau,g)=2\pi i\left(\sum_{n>0}nTr\big({X_n\over 1-X_n}P\big)
-{\rank P\over 24}\right)+O(k^{-1})&\highkflatc b\cr
X_n(\~\tau,\tau,g)\equiv (\~q^n\~P+q^nP)\Omega(g)&\highkflatc c\cr}
$$
where the result for $D$
follows from that for $\~D$ and the K-conjugation covariance of the connections
in \Kconjcov.
We note in particular that
the leading terms \highkflatc{}\ in the flat connections are functions,
so that their differential structure begins at $O(k^{-1})$.

It is instructive to check that these connections are flat, that is
$d\~D=\~dD$.
Using the high-level forms in \highkflatc{}, we know that
\eqn\hkddbar{
[D,\~D]=0
}
because the high-level connections are functions. Then,
we need only check that the high-level connections are abelian flat,
\eqn\hkabelianflat{
\del \~D=\~\del D\quad .
}
This property, and hence the flatness of the high-level connections,
follows from the identities,
\eqn\hkidentity{
q\del_q Tr\left\{\big({X_n\over 1-X_n}\big)\~P\right\}=
\~q\del_{\~q} Tr\left\{\big({X_n\over 1-X_n}\big)P\right\}=
n(\~qq)^n
Tr\left\{\~P\Omega{1\over 1-X_n}
P\Omega{1\over 1-X_n}\right\}\quad
}
which are easily checked by differentiation. We remark that the
high-level flat connections of ICFT on the sphere are also abelian-flat
\obersIII.

\subsec{High-level Bicharacters}

Having determined the high-level flat connections $\~D$ and $D$,
we may integrate eq. \caldevolution{}\ to obtain the high-level evolution
operators $\~{B}$ and ${B}$ of the flat connections,
\eqna\hkcalddbar
$$
\displaylinesno{
\~{B}(\~\tau,\tau,g)=\left(q\over \~q\right)^{\rank\~P\over 24}
\prod_{n=1}^\infty\e{
2\pi in\int_{\tau}^{\~\tau} d{\~\tau}^\prime
Tr\left\{\big({X_n({\~\tau}',\tau,g)
\over 1-X_n({\~\tau}',\tau,g)}\big)\~P\right\}
}
+O(k^{-1})&\hkcalddbar a\cr
{B}(\~\tau,\tau,g)=\left(\~q\over q\right)^{\rank P\over 24}
\prod_{n=1}^\infty\e{
2\pi in\int_{\~\tau}^\tau d\tau^\prime
Tr\left\{\big({X_n(\~\tau,\tau',g)\over 1-X_n(\~\tau,\tau',g)}\big)P\right\}
}
+O(k^{-1}) &\hkcalddbar b\cr
X_n(\~\tau,\tau,g)=
\left(\e{2\pi in\~\tau}\~P+\e{2\pi in\tau}P\right)\Omega(g)
\quad . &\hkcalddbar c\cr
}
$$
Finally, we may substitute the results \hkcalddbar{}\ into eq. \WIT{a}\ to
obtain the high-level forms of the low-spin affine-Virasoro characters,
\eqna\hkAVchar
$$
\eqalignno{
\chi(T,\~\tau,\tau,g)
&\bk{\~q}^{-{\rank\~P\over 24}}q^{-{\rank P\over 24}}
\prod_{n=1}^\infty\e{
2\pi in\int_{\tau}^{\~\tau} d{\~\tau}^\prime
Tr\left\{\big({X_n({\~\tau}',\tau,g)
\over 1-X_n({\~\tau}',\tau,g)}\big)\~P\right\}
}
{Tr(g(T))\over \Pi(\tau,\Omega(g))}\qquad &\hkAVchar a\cr
&\bk{\~q}^{-{\rank\~P\over 24}}q^{-{\rank P\over 24}}
\prod_{n=1}^\infty\e{
2\pi in\int_{\~\tau}^\tau d\tau^\prime
Tr\left\{\big({X_n(\~\tau,\tau',g)
\over 1-X_n(\~\tau,\tau',g)}\big)P\right\}
}
{Tr(g(T))\over \Pi(\~\tau,\Omega(g))}\qquad &\hkAVchar b\cr}
$$
where low spin means that the invariant Casimir of irrep $T$ is $O(k^0)$
(and so the conformal weights of irrep T are $O(k^{-1})$).
To obtain this result, we also used the high-level form of the low-spin
affine-Sugawara characters,
\eqn\hkchar{
\chi_g(T,\tau,g)\bk
q^{-{\dim g\over 24}}
{Tr(g(T))\over \Pi(\tau,\Omega(g))}
}
which follows from their explicit form in \gsol. The results \hkAVchar{}\ and
\hkchar\ are the leading terms of the high-level asymptotic expansion
of these quantities.

As a check on the high-level bicharacters \hkAVchar{}, we note
the simple intuitive result at unit source,
\eqn\hkgeql{
\chi(T,\~\tau,\tau,g=1)
\bk{\dim T\over\eta(\~\tau)^{\rank\~P}\eta(\tau)^{\rank P} }
}
where $\eta$ is the Dedekind $\eta$-function.

\newsec{When the Source is the Symmetry Group}

\subsec{Source Dependence of $h$ and $g/h$ }

We return to the case of $h$ and the $g/h$ coset constructions,
now on a general source $g\in G$, whose high-level connections and bicharacters
are included in the results above.
The answers for $h$ and $g/h$ can be obtained for any of the results of
Section 8 by choosing,
\eqn\cosetproj{
\~P=P_{g/h}={\bf 1} - P_h\quad ,\quad P=P_h
}
where $P_h$ is the projector onto $h\subset g$.

Comparing \highkflatc{}\ and \DLH{}, we see in particular that the
$h$-connection
$D(L_h,\~\tau,\tau,g)$ on a $G$ source is a function of $\~\tau$ and $\tau$,
while the $h$-connection $D(L_h,\tau,h)$ on an $H$ source is a function only
of $\tau$. Correspondingly, all the results for $h$ and $g/h$ are more
complicated for the $G$ source, and, in particular the factorization
\hexpan\ is obscured on the general source.

This is an interesting complication for $h$ and $g/h$, which should be
studied in the future.
In the present paper, we limit ourselves to understanding that the
simplification on an $H$ source is due to the  $h$-symmetry of the
K-conjugate pair $h$ and $g/h$.

When the source is restricted to $h\in H\subset G$, we may use the
$h$-invariance of the projectors,
\eqn\solvcond{
[\Omega(h),P_{g/h}]= [\Omega(h),P_{h}]=0
}
to simplify the high-level connections of $h$ and $g/h$. Then
the connections \highkflatc{}\ reduce to the forms,
\eqna\simpleform
$$
\eqalignno{
\~D(L_{g/h},\~\tau,h)&= 2\pi i\left(\sum_{n>0}n
Tr\left({\~q^n\Omega(h)\over 1-{\~q}^n\Omega(h)}P_{g/h}\right)
-{\rank P_{g/h}\over 24}\right) +O(k^{-1})&\simpleform a\cr
D(L_{h},\tau,h)&= 2\pi i\left(\sum_{n>0}n
Tr\left({q^n\Omega(h)\over 1-{q}^n\Omega(h)}P_h\right)
-{\rank P_{h}\over 24}\right) +O(k^{-1})&\simpleform b\cr}
$$
which are functions only of $\~\tau$ and $\tau$ respectively. When $H$ is
further restricted so that $G/H$ is a reductive coset space, it is easy
to check that the results \simpleform{}\ agree with the high-level
form of the exact results in eqs. \DLH{}\ and \Dbarsol.

We are now prepared to exploit this simplification in a much larger
class of ICFT's.

\subsec{The $H$-invariant CFT's}

As reviewed in Section 2,
the K-conjugate pairs $h$ and $g/h$ are only the simplest  examples
of the much larger class of ICFT's known as the $H$-invariant CFT's \Lieh,
\eqn\Iix{
\hbox{\rm ICFT}\supset\supset\hbox{\rm $H$-invariant CFT's}
\supset\supset\hbox{\rm Lie $h$-invariant CFT's}\supset\supset{\rm RCFT}\quad .
}
The space of $H$-invariant CFT's is the set of all ICFT's with a residual
global symmetry group $H$, and the $H$-symmetry, which is the symmetry group of
$\~L$ and $L$, may be a finite group or a Lie group.

The simplification seen for $h$ and $g/h$ in subsection 9.1 extends to
all the $H$-invariant CFT's. The basic point is that the bicharacters
of  any K-conjugate pair of $H$-invariant CFT's are $H$-invariant  when the
source $h$ is chosen in $H\subset G$,
\eqn\conjinv{
\chi(T,\~\tau,\tau,h_0hh_0^{-1})=
\chi(T,\~\tau,\tau,h)\quad ,\quad\forall h_0\in H\subset G
}
while a larger source breaks the $H$-symmetry. The relation \conjinv\
follows from \bichar\ and \Hinv.

At high-level on simple $g$, we can study this simplification in further
detail. According to eqs. \hkL{}\ and \Hinv, the high-level form of the
$H$-invariance of the K-conjugate pair reads,
\eqn\solvcond{
[\Omega(h),\~P]= [\Omega(h),P]=0
\quad ,\quad\forall h\in H\subset G\quad .
}
Then, choosing the source $h$ in the symmetry group $H$ of the
pair, we may use \solvcond\ in \highkflatc{}\ to obtain the flat connections
of all the $H$-invariant CFT's,
\eqna\simpleflat
$$
\eqalignno{
\~D(\~L,\~\tau,h)&=
2\pi i\left(\sum_{n>0}n
Tr\left({\~q^n\Omega(h)\over 1-{\~q}^n\Omega(h)}\~P\right)
-{\rank\~P\over 24}\right) +O(k^{-1})&\simpleflat a\cr
D(L,\tau,h)&=
2\pi i\left(\sum_{n>0}n
Tr\left({q^n\Omega(h)\over 1-{q}^n\Omega(h)}P\right)
-{\rank P\over 24}\right) +O(k^{-1})\quad .&\simpleflat b\cr}
$$
Note that, on the $H$-source, these connections are functions only of
$\~\tau$ and $\tau$ respectively.

The connections \simpleflat{}\ can be further simplified by introducing
the generalized $\Pi$-function,
\eqna\pif
$$
\displaylinesno{
\Pi(M,\tau,\Omega (h))\equiv\prod_{n=1}^{\infty}
\e{Tr\left(M\log (1-q^n\Omega(h))\right)}&\pif a\cr
\Pi(M,\tau,\Omega (h))\Pi(N,\tau,\Omega (h))=\Pi(M+N,\tau,\Omega (h))\quad ,
\quad \Pi({\bf 1},\tau,\Omega (h))\equiv \Pi(\tau,\Omega (h))
\quad\quad&\pif b\cr
\Pi(\~P{\rm\ or\ }P,\tau,\Omega (h_0hh_0^{-1}))=
\Pi(\~P{\rm\ or\ }P,\tau,\Omega (h))&\pif c\cr}
$$
where $h_0\in H$ and $\Pi(\tau,\Omega(h))$ is the simple $\Pi$-function in
\gcon{b}. Then, the connections can be written as
\eqna\simpleflatB
$$
\eqalignno{
\~D(\~L,\~\tau,h)&=
-\left(2\pi i{\rank\~P\over 24}+\~\del\log\Pi(\~P,\~\tau,\Omega (h))\right)
+O(k^{-1})&\simpleflatB a\cr
D(L,\tau,h)&=
-\left(2\pi i{\rank P\over 24}+\del\log\Pi(P,\tau,\Omega (h))\right)
+O(k^{-1})\quad .&\simpleflatB b\cr}
$$
Using this form, it is easy to obtain the evolution operators of the
flat connections,
\eqna\piMOmega
$$
\eqalignno{
\~{B}(\~\tau,\tau,h)&=\left(q\over \~q\right)^{\rank\~P\over 24}
{\Pi(\~P,\tau,\Omega(h))\over\Pi(\~P,\~\tau,\Omega(h))}+O(k^{-1})&\piMOmega
a\cr
{B}(\~\tau,\tau,h)&=\left(\~q\over q\right)^{\rank P\over 24}
{\Pi(P,\~\tau,\Omega (h))\over\Pi(P,\tau,\Omega (h))}+O(k^{-1})&\piMOmega b\cr}
$$
by integrating eq. \caldevolution{}.

Finally, we obtain the high-level, low-spin bicharacters of the $H$-invariant
CFT's,
\eqn\Liehhighk{
\chi(T,\~\tau,\tau,h)\bk
{1\over \~q^{\rank\~P\over 24}\Pi(\~P,\~\tau,\Omega(h))}
Tr(h(T))
{1\over q^{\rank P\over 24} \Pi(P,\tau,\Omega(h))}
}
from \WIT{}, \hkchar\ and \piMOmega{}, using \pif{b}\ in the form
\eqn\PplusPtwid{
\Pi(\tau,\Omega(h))=\Pi(\~P,\tau,\Omega(h))\Pi(P,\tau,\Omega(h))\quad .
}
With eq. \pif{c}, we explicitly
verify the $h$-invariance \conjinv\ of the bicharacters in \Liehhighk.

The results \piMOmega{}\ and \Liehhighk\ can also be verified directly
from eqs. \hkcalddbar{}\ and \hkAVchar{}.
Similarly, for the special case of $h$ and $g/h$ with $G/H$ a reductive
coset space, we may use the identities,
\eqn\ninei{
\Pi(P_{g/h},\~\tau,\Omega(h))=\Pi(\~\tau,\sigma(h))\quad ,\quad
\Pi(P_h,\tau,\Omega(h))=\Pi(\tau,\rho(h))
}
to check \piMOmega{}\ and \Liehhighk\ against the high-level forms of
the exact results in \caldeqOM\ and \Dcalomeg.

\subsec{Candidate Characters for the Lie $h$-invariant CFT's}

To obtain the characters of the individual ICFT's, it is necessary to
factorize the biconformal characters,
\eqn\charexpand{
\chi(T,\~\tau,\tau,h)=
\sum_{\nu}\chi_{\~L}^\nu(T,\~\tau,h)\chi_{L}^{\vphantom{g}\nu}(T,\tau,h)
}
into the conformal characters $\chi_{\~L}^{\vphantom\nu}$
and $\chi_L^{\vphantom g}$ of the $\~L$ and the $L$ theory respectively.
As on the sphere \obersII, there are many factorizations, or bases, of the form
\charexpand, but we are interested only in those factorizations for which
the conformal characters exhibit modular covariance. See Ref. \obersII\ for an
analogous factorization of the bicorrelators of ICFT on the sphere,
in which the conformal
correlators of $\~L$ and $L$ are covariant under the braid group.

Here, we make a modest beginning in this direction, obtaining
high-level candidate characters for the Lie $h$-invariant CFT's \Lieh,
which form the subset of all H-invariant CFT's with H a Lie group.
This class of ICFT includes $h$ and
the $g/h$ coset constructions as a small subspace. For all these theories,
We know from \TadjL\ and \conjinv\ that,
\eqna\TadjLB
$$
\displaylinesno{
[T_A^{\rm adj},\~L]= [T_A^{\rm adj},L]=0
\quad ,\quad A=1,\ldots ,\dim h &\TadjLB a\cr
(\=E_A+E_A)\chi(T,\~\tau,\tau,h)=0 &\TadjLB b\cr}
$$
and we may hope to follow the intuition gained from $h$ and the $g/h$ coset
constructions.

More precisely, we restrict ourselves to the Lie $h$-invariant CFT's
with simple $h\subset g$. Then we know \Lieh\ that one of the theories, say
$\~L$, has a local Lie $h$-invariance (like $L_{g/h}$)
\eqn\nineII{
[J_A(m),\~L(n)]=0\quad ,\quad m,n\in\IZ
}
while its K-conjugate partner (like $L_h$) carries only the
corresponding global invariance,
\eqn\nineIII{
[J_A(0),L(n)]=0\quad ,\quad m\in\IZ\quad .
}
In this case, as for $h$ and $g/h$, we may adopt as a working hypothesis
that all the source dependence of the bicharacters is associated to the
global theory.

At high-level on simple $g$, the low-spin bicharacters of the Lie $h$-invariant
CFT's are given by the result \Liehhighk\ with
\eqn\nineIV{
[T_A^{\rm adj},\~P]= [T_A^{\rm adj},P]=0
}
so each factor in the high-level bicharacters of the Lie $h$-invariant CFT's
on the $H$ source are explicitly $h$-invariant.
Then we may $h$-character expand the local theory $\~L$ in \Liehhighk\ to
obtain the factorized bicharacters,
\eqn\charexpandB{
\chi(T,\~\tau,\tau,h)\bk
\sum_{T^h}\chi_{\~L}^{\vphantom\nu}(T,T^h,\~\tau)
\chi_{L}^{\vphantom g}(T,T^h,\tau,h)
}
where the sum is over all unitary irreps $T^h$ of $h$, and
\eqna\chiLLtilde
$$
\eqalignno{
\chi_{\~L}^{\phantom g}(T,T^h,\~\tau)&\bk
\int dh
{Tr(h^*(T^h))Tr(h(T))
\over
\~q^{\rank\~P\over 24}
\Pi(\~P,\~\tau,\Omega(h))}&\chiLLtilde a\cr
& & \cr
\chi_L^{\phantom g}(T^h,\tau,h)&\bk
{Tr(h(T^h))\over
q^{\rank P\over 24}
\Pi(P,\tau,\Omega(h))}&\chiLLtilde b\cr}
$$
are the high-level candidate characters for the Lie $h$-invariant CFT's.

As a check on the candidate characters \chiLLtilde{}, we reconsider the
simple case of $h$ and
the $g/h$ coset constructions, with $G/H$ a reductive coset space.
In this case, the candidate characters reduce to the
high-level characters of $h$ and $g/h$,
\eqna\hkcosetcon
$$
\eqalignno{
\chi_{L_{g/h}}^{\phantom g}(T,T^h,\~\tau)&\bk
\int dh
{Tr(h^*(T^h)) Tr(h(T))
\over
\~q^{{\rm dim}(g/h)\over 24}
\Pi(\~\tau,\sigma(h))}&\hkcosetcon a\cr
& & \cr
\chi_{L_h}^{\phantom g}(T^h,\tau,h)&\bk
{Tr(h(T^h))\over
q^{{\rm dim}h\over 24}
\Pi(\tau,\rho(h))}&\hkcosetcon b\cr}
$$
which agree with the high-level forms of the exact results
in \cosetchar{}\ and \hathchar{}.

The next step is to test the candidate characters for modular covariance,
or to further decompose the candidates until modular covariance is obtained.
This investigation is beyond the scope of the present paper, but
we may set the stage with some simple remarks.

We know that the modular transformation $\tau\to -{1\over\tau}$ mixes low
spin with all spin, and we have checked in examples that, at high level, this
transformation is dominated by high spin (of order the level for $SU(2)$).
Thus, high-spin candidate characters are also needed to study modular
covariance
at high level.
Because all representation dependence of the bicharacters comes from the
affine-Sugawara characters $\chi_g(T)$, such high-spin candidate characters
for the Lie-$h$ invariant ICFT's can be obtained as above, from the
high-level form of the high-spin affine-Sugawara characters.
\nref\Tsey{A.A Tseytlin, {\it ``On a `Universal' Class of WZW-Type conformal
Models''}, CERN-TH.7068/93, hep-th/9311062, 1993.}
\nref\dBCH{J. de Boer, K. Clubok and M.B. Halpern, {\it Linearized Form of the
Generic Affine-Virasoro Action}, UCB-PTH-93/34, LBL-34938, ITP-SB-93-88,
hep-th/9312094.}

Although this program is technically involved, it is expected
that chiral modular covariant characters and non-chiral modular invariants
exist in ICFT, just as braid-covariant correlators have been found on the
sphere \obersII. This expectation has further support in the case of the
high-level smooth ICFT's studied here, because diffeomorphism-invariant
world-sheet actions \refs{\HYII,\Tsey,\dBCH}\ are known for the generic
theory of this type.

\newsec{A Geometric Formulation}

The characters studied in the sections above were defined with a
conventional Lie source, but we wish to point out the geometric form that
our problem takes on an affine source $\^\gamma$,
in the centrally-extended loop group $\^L G$ of affine $g$.

We write the affine source as
\eqn\ydecomp{
\^\gamma(x,y)=\e{iy\^k}\^g(x)
}
where $y$ and $x^{\alpha\mu}$, $\alpha=1,\ldots ,\dim g,\ \mu\in\IZ$ are
the coordinates on the loop group manifold and $\^k$ is the level operator,
or central element. The $y$-independent
factor $\^g$ can be chosen in many bases such as,
\eqn\expmap{
\^g(x)=\exp( i\sum_{am}\beta^{am}(x)J_a(m) )
}
or normal-ordered forms such as the Borel decompositions in Refs.
\ref\BF{D. Bernard and G. Felder, Commun. Math. Phys. {\bf 127} (1990) 145.}\
and \ref\wati{W. Taylor, Berkeley PhD Thesis {\it Coadjoint Orbits and
Conformal Field Theory}, 1993, UCB-PTH-93/26, LBL-34507, hep-th/9310040 .}.
In practice, one may wish to choose a basis of $\^g$ which simplifies the
Laplacians in the formulation below.

Define the affine-Virasoro characters on the affine source as
\eqn\affsource{
\chi(T,\~\tau,\tau,\^\gamma)=
Tr_\T\big(\~q^{\~L(0)-\~c/24}q^{L(0)-c/24}\^\gamma\big)\quad .
}
Then, following the development in the earlier sections, we introduce
left and right invariant vielbeins,
inverse vielbeins and affine Lie derivatives on the loop group as follows,
\eqna\affviel
$$
\displaylinesno{
e_\Lambda=-i\^\gamma^{-1}\del_\Lambda\^\gamma=
e_\Lambda^{\ L}{\cal J}_L
\quad ,\quad
\CE_L=-ie_L^{\ \Lambda}\del_\Lambda
\quad ,\quad
\CE_L\^\gamma=\^\gamma {\cal J}_L &\affviel a\cr
\=e_\Lambda=-i\^\gamma\del_\Lambda\^\gamma^{-1}=
\=e_\Lambda^{\ L}{\cal J}_L
\quad ,\quad
\=\CE_L=-i\=e_L^{\ \Lambda}\del_\Lambda
\quad ,\quad
\=\CE_L\^\gamma=-{\cal J}_L\^\gamma &\affviel b\cr
{\cal J}_L=(J_a(m),\^k)\quad ,\quad
\CE_L=(\CE_a(m),\CE_y)
\quad ,\quad
\=\CE_L=(\=\CE_a(m),\=\CE_y)\quad
&\affviel c\cr
\Lambda=(\alpha\mu,y)\quad ,\quad L=(am,y)\quad
&\affviel d\cr}
$$
where $[\CJ_L,\CJ_M]=if_{LM}^{\ \ N}\CJ_N$ is the affine algebra and the
vielbeins $e_\Lambda^{\ L}$, $\=e_\Lambda^{\ L}$ and inverse vielbeins
$e_L^{\ \Lambda}$, $\=e_L^{\ \Lambda}$
are independent of the operators $\CJ_L$. The affine Lie derivatives
$\CE_L$ and $\=\CE_L$ in \affviel{}\ satisfy two commuting affine algebras
with central elements $\CE_y$ and $\=\CE_y$ respectively.

With these tools, it is straightforward to see that the bicharacters
\affsource\
satisfy the heat-like equations,
\eqna\looplaplace
$$
\displaylinesno{
\~\del\chi =\~D\chi\quad ,\quad \del\chi =D\chi
&\looplaplace a\cr
\~D(\^\gamma) =-2\pi i\~\Delta(\^\gamma)=2\pi i\~L^{ab}\big(\CE_a(0)\CE_b(0)
+2\sum_{m>0}\CE_a(-m)\CE_b(m)\big)&\looplaplace b\cr
D(\^\gamma) =-2\pi i\Delta(\^\gamma)=2\pi iL^{ab}\big(\CE_a(0)\CE_b(0)
+2\sum_{m>0}\CE_a(-m)\CE_b(m)\big)&\looplaplace c\cr}
$$
and the usual affine-Sugawara boundary condition,
\eqna\looplaplaceB
$$
\displaylinesno{
\chi(T,\tau,\tau,\^\gamma)=\chi_g(T,\tau,\^\gamma)=
Tr_\T\big(q^{L_g(0)-c_g/24}\^\gamma\big) &\looplaplaceB a\cr
\del\chi_g=D_g\chi_g\quad ,\quad D_g=\~D+D &\looplaplaceB b\cr
D_g(\^\gamma)=-2\pi i\Delta_g(\^\gamma)=2\pi i L^{ab}_g\big(\CE_a(0)\CE_b(0)
+2\sum_{m>0}\CE_a(-m)\CE_b(m)\big)&\looplaplaceB c\cr}
$$
where $\chi_g(T,\tau,\^\gamma)$ are the affine-Sugawara characters on the
affine
source. The bicharacters and the affine-Sugawara characters also satisfy an
analogous heat-like system with $\CE_a(m)\to\=\CE_a(m)$.

The objects $\~\Delta$, $\Delta$ and $\Delta_g$
are three mutually-commuting Laplacians on the
centrally-extended loop group. It is easy to verify that the affine-Virasoro
connections $\~D$, $D$ are flat,
$\~\del D+D\~D=\del\~D+\~D D$, and moreover,
\eqn\afflat{
\~\del D=\del\~D=0\quad ,\quad
[\~D,D]=0\quad
}
so the connections are also abelian flat.

In further detail, we find from \ydecomp\ that
\eqna\vielproperties
$$
\displaylinesno{
e_{\alpha\mu}=-i\^g^{-1}\del_{\alpha\mu}\^g=e_{\alpha\mu}^{\ \ am}J_a(m)
+e_{\alpha\mu}^{\ \ y}\^k &\vielproperties a\cr
e_y^{\ L}=\delta_y^{\ L}\quad ,\quad e_{\alpha\mu}^{\ \ L}\
\hbox{\rm is independent of $y$} &\vielproperties b\cr
e_y^{\ \Lambda}=\delta_y^{\ \Lambda}\quad ,\quad e_{am}^{\ \ \Lambda}\
\hbox{\rm is independent of $y$} &\vielproperties c\cr
e_{am}^{\ \ \alpha\mu}e_{\alpha\mu}^{\ \ bn}=
\delta_{am}^{\ \ bn}\quad ,\quad
e_{\alpha\mu}^{\ \ am}e_{am}^{\ \ \beta\nu}=
\delta_{\alpha\mu}^{\ \ \beta\nu}\quad ,\quad
e_{am}^{\ \ \ y}=
-e_{am}^{\ \ \alpha\mu}e_{\alpha\mu}^{\ \ \ y}
&\vielproperties d\cr}
$$
and similarly for the $\=e$'s. This gives the explicit forms for the left and
right invariant Lie derivatives,
\eqna\aflied
$$
\displaylinesno{
\CE_y=-i\del_y\quad ,\quad
\CE_a(m)=-i\big(e_{am}^{\ \ \alpha\mu}\del_{\alpha\mu}
+e_{am}^{\ \ y}\del_y\big)
&\aflied a\cr
\=\CE_y=i\del_y\quad ,\quad
\=\CE_a(m)=-i\big(\=e_{am}^{\ \ \alpha\mu}\del_{\alpha\mu}
+\=e_{am}^{\ \ y}\del_y\big)
&\aflied b\cr}
$$
where the sign difference of $\CE_y$ and $\=\CE_y$ comes from
$\=e_y^{\ \Lambda}=-\delta_y^{\ \Lambda}$.

We are primarily interested in the reduced
bicharacters $\chi(\^g)$, which satisfy
\eqna\reducedchar
$$
\displaylinesno{
\chi(T,\~\tau,\tau,\^\gamma)=\e{iyk}\chi(T,\~\tau,\tau,\^g)
&\reducedchar a\cr
\chi(T,\~\tau,\tau,\^g)=
Tr_\T\big(\~q^{\~L(0)-\~c/24}q^{L(0)-c/24}\^g\big)
&\reducedchar b\cr}
$$
where $\^k$ is replaced by the level $k$ in the reduced quantities.
It follows from \looplaplace{}\ and \reducedchar{a}\
that the reduced bicharacters satisfy the heat-like system,
\eqna\looplaplaceC
$$
\displaylinesno{
\~\del\chi(\^g)=\~D(\^g)\chi(\^g)\quad ,\quad
\del\chi(\^g) =D(\^g)\chi(\^g)
&\looplaplaceC a\cr
\~D(\^g) =-2\pi i\~\Delta(\^g)=2\pi i\~L^{ab}\big(E_a(0)E_b(0)
+2\sum_{m>0}E_a(-m)E_b(m)\big)&\looplaplaceC b\cr
D(\^g) =-2\pi i\Delta(\^g)=2\pi iL^{ab}\big(E_a(0)E_b(0)
+2\sum_{m>0}E_a(-m)E_b(m)\big)&\looplaplaceC c\cr}
$$
where the reduced affine Lie derivatives $E_a(m)$ and $\=E_a(m)$ are
\eqna\redlie
$$
\eqalignno{
E_a(m)&=-ie_{am}^{\ \ \alpha\mu}\del_{\alpha\mu}
+ke_{am}^{\ \ y}=
-ie_{am}^{\ \ \alpha\mu}\big(\del_{\alpha\mu}-ike_{\alpha\mu}^{\ \ y}\big)
&\redlie a\cr
\=E_a(m)&=-i\=e_{am}^{\ \ \alpha\mu}\del_{\alpha\mu}
+k\=e_{am}^{\ \ y}=
-i\=e_{am}^{\ \ \alpha\mu}\big(\del_{\alpha\mu}+ik\=e_{\alpha\mu}^{\ \ y}\big)
\quad . &\redlie b\cr}
$$
These differential operators satisfy
\eqna\EEbar
$$
\displaylinesno{
E_a(m)\^g=\^gJ_a(m)\quad ,\quad \=E_a(m)\^g=-J_a(m)\^g &\EEbar a\cr
[E_a(m),E_b(n)]=if_{ab}^{\ \ c}E_c(m+n)+mk\eta_{ab}\delta_{m+n,0}
&\EEbar b\cr
[\=E_a(m),\=E_b(n)]=if_{ab}^{\ \ c}\=E_c(m+n)-mk\eta_{ab}\delta_{m+n,0}
&\EEbar c\cr
[E_a(m),\=E_b(n)]=0
&\EEbar d\cr}
$$
and we remark that the the left and right invariant operators satisfy the
affine algebra of $g$ at level $k$ and $-k$ respectively. It follows that
$Q_a(m)\equiv E_a(m)+\=E_a(m)$ satisfies the loop algebra of $g$.

We emphasize that the result \EEbar{b}\ is a class of new representations
of the affine algebra, one for each basis choice of $\^g$. An example of this
system, for a particular
basis of affine $su(2)$, has been studied in Ref. \ref\ANS{V. Aldaya and
J. Navarro-Salas, Commun. Math. Phys. {\bf 113} (1987) 375.}.
As an example on all affine $g$, one may choose the basis
\eqn\ourbasis{
\^g(x)=\exp\big(ix^{\alpha\mu}e_{\alpha\mu}^{\ \ am}(0)J_a(m)\big)\quad .
}
Then we obtain the explicit forms of the left invariant quantities,
\eqna\explicitE
$$
\displaylinesno{
e_{am}^{\ \ \alpha\mu}(x)=M(x)_{am}^{\ \ bn}e_{bn}^{\ \ \alpha\mu}(0)
\quad ,\quad e_{am}^{\ \ \ y}(x)=M(x)_{am}^{\ \ \ y}
&\explicitE a\cr
e_{\alpha\mu}^{\ \ \ y}(x)=
e_{\alpha\mu}^{\ \ am}(0)M^{-1}(x)_{am}^{\ \ \ y}
&\explicitE b\cr
M(x)={\log\^g(\^T^{adj},x)\over \^g(\^T^{adj},x)-1}\quad ,\quad
\^g(\^T^{adj},x)=\exp\big(ix^{\alpha\mu}e_{\alpha\mu}^{\ \ am}(0)
\^T_{am}^{adj}\big)
&\explicitE c\cr
(\^T_{am}^{adj})_{bn}^{\ \ cr}=\delta_{m+n,r}(T_a^{adj})_b^{\ c}\quad ,\quad
(\^T_{am}^{adj})_{bn}^{\ \ \ y}=n\eta_{ab}\delta_{m+n,0}
&\explicitE d\cr}
$$
and similarly for the right invariant quantities,
where the non-zero elements of the affine adjoint matrix
$(\^T_L^{adj})_M^{\ \ N}=-if_{LM}^{\ \ \ N}$ are given in \explicitE{d}.

The reduced system \looplaplaceC{}, taken with the usual affine-Sugawara
boundary condition,
\eqn\bc{
\chi_g(T,\tau,\^g)=\chi(T,\tau,\tau,\^g)\quad ,\quad
\del\chi_g(\^g)=D_g(\^g)\chi_g\quad ,\quad
D_g(\^g)=\~D(\^g)+D(\^g)=-2\pi i\Delta_g(\^g)\quad
}
can be further analyzed using the machinery developed in the earlier sections.
In particular, the reduced objects $\~\Delta(\^g)$, $\Delta(\^g)$ and
$\Delta_g(\^g)$, which represent $-\~L(0)$, $-L(0)$ and $L_g(0)$ respectively,
are three mutually commuting generalized Laplacians on the centrally-extended
loop group, and the reduced connections $\~D$, $D$ are flat and abelian flat.
Moreover, the reduced bicharacters $\chi(\^g)$ are
uniquely determined from eqs. \WIT{}\ and \caldevolution{},
\eqn\heatkernel{
\chi(T,\~\tau,\tau,\^g)=\e{-2\pi i(\~\tau-\tau)\~\Delta(\^g)}\chi_g(T,\tau,\^g)
=\e{-2\pi i(\tau-\~\tau)\Delta(\^g)}\chi_g(T,\~\tau,\^g)
}
in terms of the reduced affine-Sugawara characters \bc, whose explicit
form we will not obtain in this paper.

The form of the solution \heatkernel\ is seen more clearly by introducing
the simultaneous eigenvectors $\psi_n(T,\^g)$ of the three Laplacians,
\eqna\eigen
$$
\displaylinesno{
-\~\Delta\psi_n=\~\lambda_n\psi_n\quad ,\quad
-\Delta\psi_n=\lambda_n\psi_n\quad ,\quad
-\Delta_g\psi_n=\lambda_n^g\psi_n &\eigen a \cr
\~\lambda_n+\lambda_n=\lambda_n^g\quad &\eigen b \cr}
$$
in the Hilbert space of affine irrep $T$, where the eigenvalues
$\~\lambda_n(T)$, $\lambda_n(T)$ and $\lambda_n^g(T)$ are the conformal
weights of the states $\psi_n(T,\^g)$ under the stress tensors $\~T$, $T$ and
$T_g$. The basis \eigen{}\ is therefore the simultaneous $L$-basis
(see Section 3) for all levels of the affine irrep $T$.
For unitary theories, $\~L(0)$, $L(0)$ and $L_g(0)$ are hermitean in an inner
product with a non-negative norm, so, in the representation above, there is
an induced inner product $\langle A|B\rangle$ in which the Laplacians are
hermitean and the eigenvectors are orthonormal
$\langle\psi_m(T)|\psi_n(T)\rangle =\delta_{m,n}$.

Using these eigenvectors, we find the unique solution of the heat-like system,
\eqna\unique
$$
\displaylinesno{
\chi(T,\~\tau,\tau,\^g)=\sum_n\left({\~q\over q_0}\right)^{\~\lambda_n(T)}
\left({q\over q_0}\right)^{\lambda_n(T)}\psi_n(T,\^g)
\langle\psi_n(T)|\chi_g(T,\tau_0)\rangle  &\unique a\cr
\chi_g(T,\tau,\^g)=\chi(T,\tau,\tau,\^g)=
\sum_n\left({q\over q_0}\right)^{\lambda_n^g(T)}\psi_n(T,\^g)
\langle\psi_n(T)|\chi_g(T,\tau_0)\rangle  &\unique b\cr}
$$
where $q_0=\exp(2\pi i\tau_0)$ is a regular reference point.

The solution \unique{a}\ for the bicharacters follows from eq. \heatkernel,
using the expansion \unique{b}\ of the affine-Sugawara characters, and it is
easy to check that the full solution \unique{a,b}\ solves the heat-like
equations \looplaplaceC{a}\ and the affine-Sugawara boundary condition \bc.
Moreover, the bicharacters
and the affine-Sugawara characters are independent of the reference point,
as they should be. For example, one finds
\eqn\refpoint{
\del_{\tau_0}\chi=
\sum_n\left({\~q\over q_0}\right)^{\~\lambda_n}
\left({q\over q_0}\right)^{\lambda_n}\psi_n
\langle\psi_n|
\big(-2\pi i\lambda_n^g+D_g(T,\tau_0,\^g)\big)\chi_g(T,\tau_0,\^g)\rangle=0
}
where hermiticity of $D_g$ is used in the last step.

The bicharacters of the earlier sections can be obtained from these
bicharacters by restricting the affine source to a Lie source.
The advantage of the geometric formulation is that we now have the flat
connections in closed form, which may be useful in the investigation of
global properties such as factorization and modular covariance.

\newsec{Conclusions}

Irrational conformal field theory (ICFT) includes rational conformal field
theory as a small subspace. So far, the only known path into the space of
ICFT's is the general affine-Virasoro construction \refs{\HK,\Moroz} on the
currents of
affine Lie $g$. Recently, dynamical equations for the correlators of ICFT have
been obtained on the sphere \refs{\obersI,\obersII}, where they are understood
as generalized
Knizhnik-Zamolodchikov equations with flat connections \obersIII.

In this paper we have begun the study of ICFT on the torus, following the
paradigm on the sphere. In particular, we have defined the affine-Virasoro
characters, or bicharacters, which involve the two commuting Virasoro
operators of any
affine-Virasoro construction, and we have shown that the bicharacters satisfy
heat-like equations with flat connections.

As a first example of the formulation, we have solved the system
completely for the simple case of $h$ and the $g/h$ coset constructions,
obtaining a new integral representation for the general coset characters.

In a second application, we have solved for the high-level bicharacters
of the general ICFT on simple $g$, and proposed a set of high-level candidate
characters for the Lie $h$-invariant CFTs \Lieh, which is the set of all ICFTs
with a Lie Symmetry.

A next step is to analyze the high-level candidate characters with respect to
the modular group. For this investigation, one should begin with some of the
many explicit examples of Lie-$h$ invariant CFT's, beyond the coset
constructions. We did not attempt to factorize the general theory on a
source larger then its symmetry group, which is an important problem
because the generic ICFT has no residual symmetry. As a first step in this
direction, one should study the factorization of $h$ and $g/h$ when
the source is larger then $H$.

Finally, we noted a geometric formulation of the system on an affine source,
where the flat connections are generalized Laplacians on the centrally-extended
loop group. These Laplacians involve new first-order
differential representations of affine Lie algebra.

\vskip1cm
\centerline{\bf Acknowledgments}

We thank I. Bars, D. Bernard, E. Kiritsis, W. Taylor and S. Yankielowicz
for helpful remarks, and we are indebted to H. Ooguri for a discussion which
stimulated this investigation.

\acknowledge\

\listrefs

\bye